\documentclass[preprint2]{aastex}

\slugcomment{Not to appear in Nonlearned J., 45.}

\shorttitle{Proper Motion of 3C 66B}
\shortauthors{Sudou and Iguchi}

\begin{document}


\title{Proper Motion of the Sub-Parsec Scale Jet in the Radio Galaxy 3C 66B}


\author{Hiroshi Sudou}
\affil{Faculty of Engineering, Gifu University, 1-1 Yanagido, Gifu, Gifu 501-1193, Japan; sudou@gifu-u.ac.jp}

\and

\author{Satoru Iguchi}
\affil{National Astronomical Observatory of Japan, 2-2-1 Osawa, Mitaka, Tokyo 181-8588, Japan; s.iguchi@nao.ac.jp}







\begin{abstract}

We present proper motion of the sub-parsec scale jet in a nearby elliptical galaxy 3C 66B. 
Observations were made using the VLBA and partly Effelsburg 100-m telescope at 2.3 GHz and 8.4 GHz at 10 epochs over 4 years. 
The 8.4 GHz images showed that a proper motion increases from 0.21 to 0.70 mas year$^{-1}$, corresponding to an apparent speed of  0.30 $c$ to 0.96 $c$, with a distance from the core on a sub-parsec scale. 
Our investigation suggests that the apparent increase of the proper motion can be explained by changes in the viewing angle, according to a relativistic beaming model. 
However, we still cannot eliminate the possibility that acceleration of the jet outflow speed or of changes of emissivity profile in the two-zone jet might be found in 3C 66B.

\end{abstract}





\section{Introduction}

3C 66B is a giant elliptical galaxy with a redshift ($z$) of 0.0213 \citep{huchra99} within the cluster A347 \citep{fanti82} and the Perseus super cluster \citep{gregory81}.
It has a prominent face-on dust lane, which indicates the occurrence of galaxy merger events \citep{verdoes99}.
This source is also known as an FR I radio galaxy and it has a strong jet and counterjet that extend to about 100 kiloparsecs (kpc), 
which have been observed at radio \citep{hardcastle96}, infrared \citep{tansley00},
optical \citep{macchetto91},  and X-ray waves \citep{hardcastle01}. 

High-resolution imaging using very long baseline interferometry (VLBI) has revealed an asymmetric core-jet structure similar to that observed in radio galaxies \citep{fraix97, xu00, giovannini01}. 
Observation at 5.0 GHz reveals the presence of a faint counterjet \citep{giovannini01}. The intensity ratio between the jet and counterjet was estimated to be approximately 10 within a distance of 2 mas from the core, 
indicating possible acceleration of the jet outflow on a parsec scale. 
The orbital motion of the compact radio core in 3C 66B, which has a period of 1.05 $\pm$ 0.03 years, was observed with a position accuracy of 10 micro arcseconds ($\mu$as) using phase-referencing
VLBI \citep{sudou03}. 

Here, we report the detection of the proper motion of the sub-parsec scale jet in 3C 66B based on multi-epoch observation over 4 years. 
Observations and data reduction are  described in \S 2, and measurements of proper motion, including the calibration of the systematic shift of the position of the core are presented in \S 3.
\S 4 focuses on the increase of the apparent velocity  of the jet with distance from the core. 

In this paper, we use a Hubble constant ($H_0$) of 71 km s$^{-1}$ Mpc$^{-1}$ , a matter density ($\Omega_{\rm M}$) of 0.27, and a vacuum energy ($\Omega_{\Lambda}$) of 0.73, which results in
an angular size or separation of 1 milliarcsecond (mas) corresponding to 0.436 pc at the distance of 3C 66B.

\section{Observations and Data Reduction}

3C 66B was observed at 2.3 GHz and 8.4 GHz using VLBI during the period of 2001.20 to 2002.45 (6 epochs) and of 2004.80 to 2005.69 (5 epochs). 
In this paper, we name these two periods Period 1 and Period 2. Participating antennas for each epoch are listed in Table 1. 
We used the NRAO\footnote{The National Radio Astronomy Observatory is a facility of the National Science Foundation operated under cooperative agreement by Associated Universities, Inc.} VLBA \citep{napier94} for the first 10 epochs and the VLBA  and Effelsberg 100-m telescope of the MPIfR for the last epoch. 
Due to bad weather conditions and antenna problems, the 8.4 GHz data for the second epoch (2001.48) showed very low quality in reproducing the faint jet structure, 
and so we removed this data from further imaging analysis. 
In addition, we only observed at 8.4 GHz for the last epoch (2005.69). Consequently, we obtained 10 sets of data for each frequency. 

The data were recorded in the VLBA format with 2 intermediate frequency bands with a bandwidth of 8 MHz each for each frequency. 
The total bit rate for both frequencies was 128 Mbps using a 2-bit sampling mode.
The observed data were correlated with the VLBA correlator in Socorro.
The correlated data for 3C 66B were calibrated using the AIPS package developed by NRAO.
The data for all epochs were treated in a similar fashion;  the residual phase delays and delay rates that were mainly due to antenna position errors and atmospheric phase fluctuations were calibrated using global fringe-fitting \citep{schwab83}.
Visibility amplitudes were calibrated using measured system temperature and antenna gain curves. 
Bandpass calibration was performed using 0133+476 for all epochs. 

The calibrated visibility data were coherently averaged over 10 -- 30 seconds according to data quality, which varied between epochs due to the different weather conditions.  
The images were made with DIFMAP \citep{shepherd97} using CLEAN and self-calibration.  

\begin{figure}
  \begin{center}
 \epsscale{0.75}
\plotone{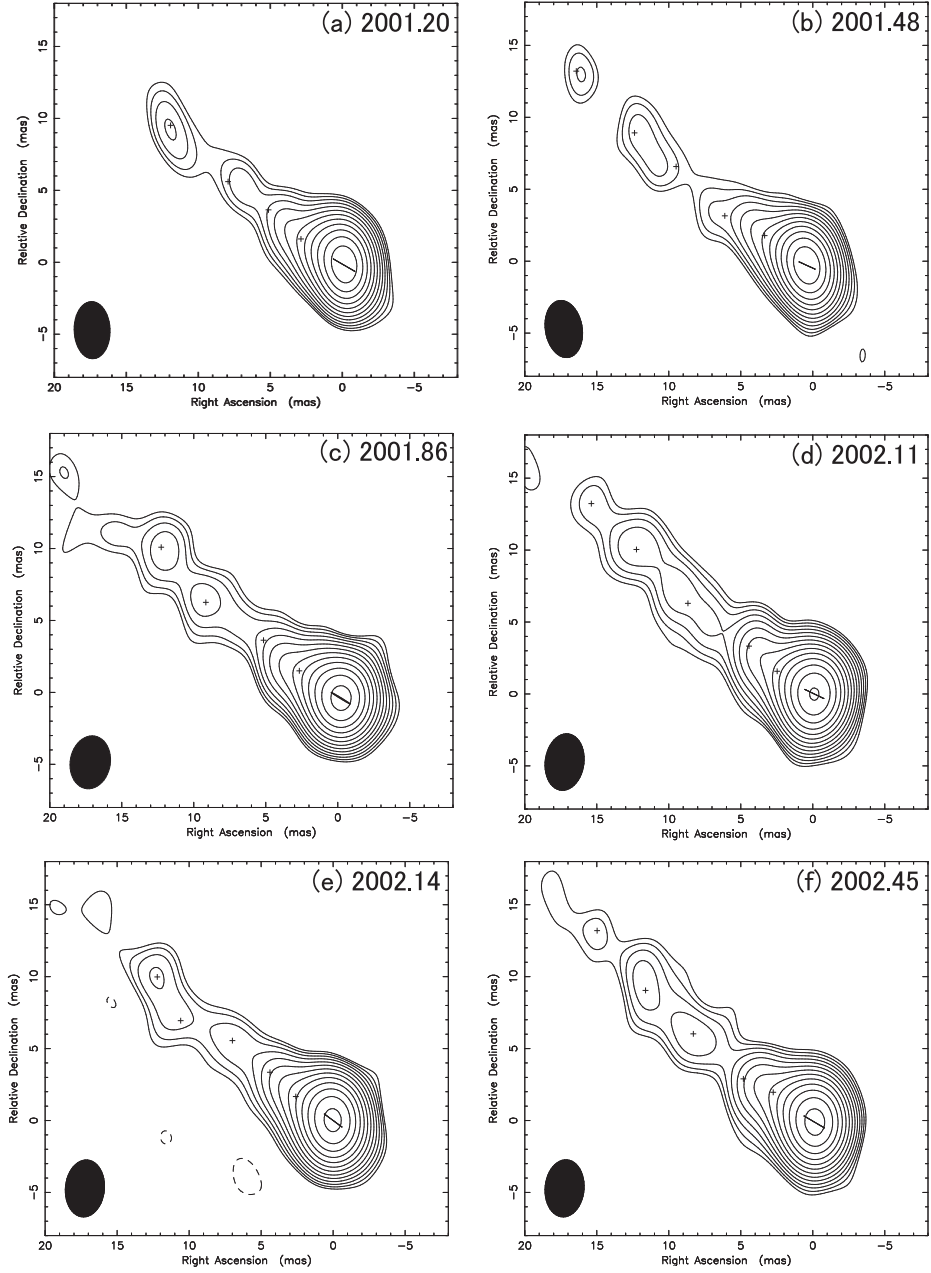}
  \caption{Six 2.3 GHz images of 3C 66B during 2001.20 to 2002.45 (Period 1). Contours are drawn at 3-$\sigma \times (1, \sqrt{2},2,2\sqrt{2},...$). Each value of $\sigma$ is 0.5, 0.6, 0.6, 0.4, 0.6 and 0.5 mJy beam$^{-1}$ from (a) to (f), respectively.
The synthesized beam is shown at the bottom-left of the image, the  accurate size of which is also shown in Table 1. The cross marks indicate the fitted position  of knots using a delta functionin.  The Gaussian component of the core is also shown.
\label{fig:smap1}
}
 \end{center}
\end{figure}

\begin{figure}
  \begin{center}
 \epsscale{0.75}
 \plotone{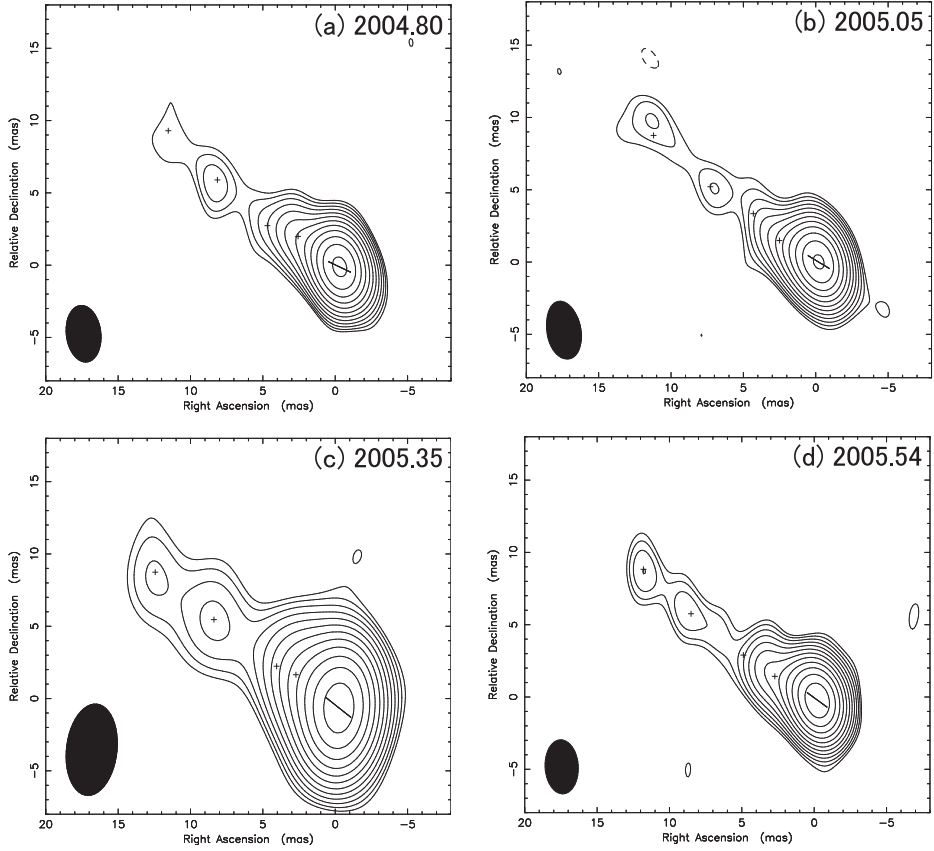}   \caption{Four 2.3 GHz images of 3C66B during 2004.80 to 2005.54 (Period 2). Contours are drawn at 3-$\sigma \times (1, \sqrt{2},2,2\sqrt{2},...$). Each value of $\sigma$ is 0.7 mJy beam$^{-1}$  for these epochs.  
The synthesized beam is shown at the bottom-left of the image, the  accurate size of which is also shown in Table 1. The cross marks indicate the fitted position  of knots using a delta functionin.  The Gaussian component of the core is also shown.
\label{fig:smap2}
}  
\end{center}
\end{figure}

\begin{figure}
  \begin{center}
 \epsscale{0.75}
 \plotone{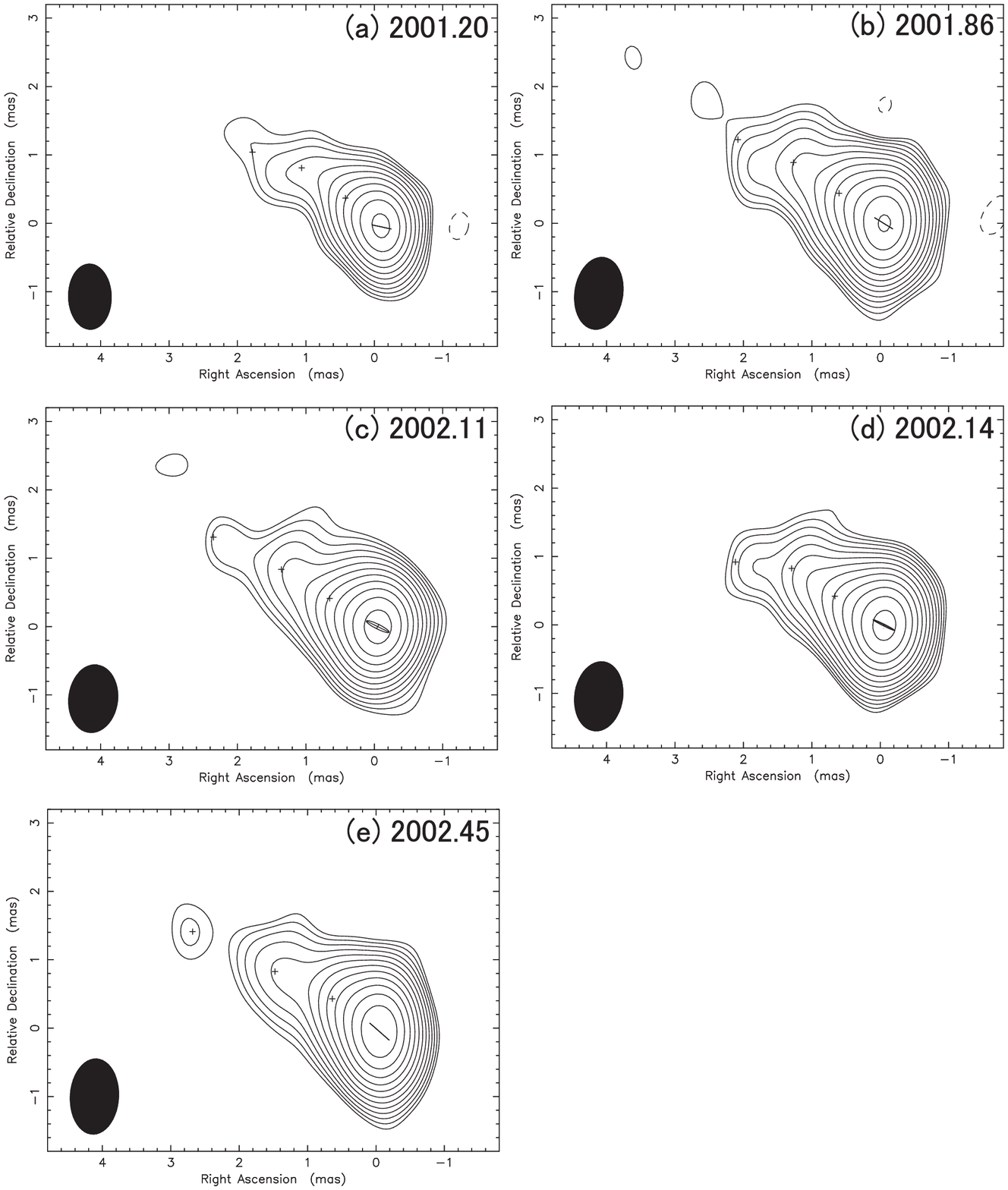}
  \caption{Five 8.4 GHz images of 3C66B during 2001.20 to 2002.45 (Period 1). Contours are drawn at 3-$\sigma \times (1, \sqrt{2},2,2\sqrt{2},...$). Each value of $\sigma$ is 0.8, 0.8, 0.9, 0.6 and 0.7 mJy beam$^{-1}$  from (a) to (e), respectively.
The synthesized beam is shown at the bottom-left of the image, the  accurate size of which is also shown in Table 1. The cross marks indicate the fitted position  of knots using delta functionin.  The Gaussian component of the core is also shown.
}
\label{fig:xmap1}
 \end{center}
\end{figure}

\begin{figure}
  \begin{center}
 \epsscale{0.75}
 \plotone{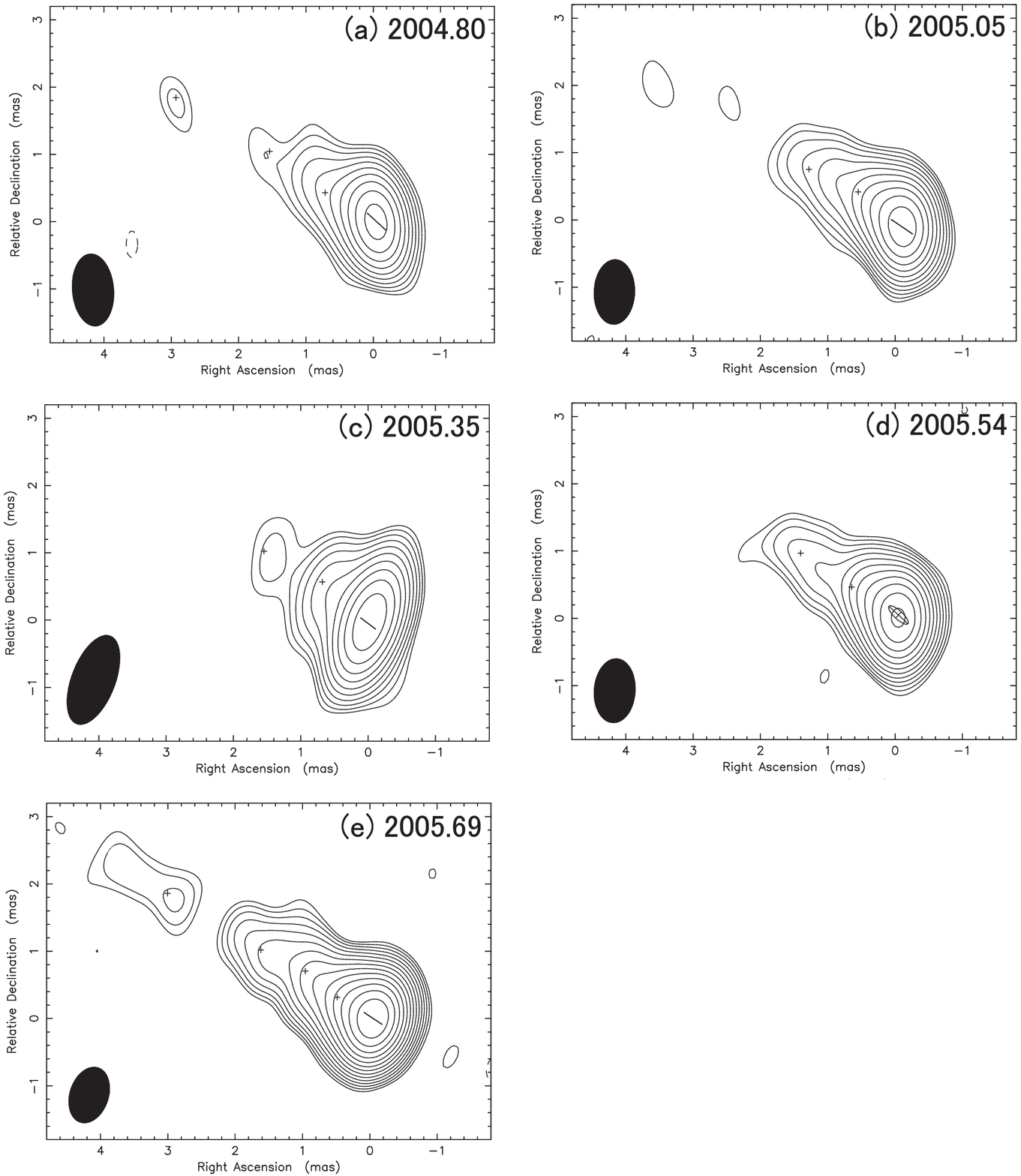}
  \caption{Five 8.4 GHz images of 3C66B during 2004.80 to 2005.69 (Period 2). Contours are drawn at 3-$\sigma \times (1, \sqrt{2},2,2\sqrt{2},...$). Each value of $\sigma$ is 1.2, 0.7, 1.0, 0.8 and 0.4 mJy beam$^{-1}$  from (a) to (e), respectively.
The synthesized beam is shown at the bottom-left of the image, the  accurate size of which is also shown in Table 1. The cross marks indicate the fitted position  of knots using a delta functionin. The Gaussian component of the core is also shown.
\label{fig:xmap2}
}
\end{center}
\end{figure}

Figures \ref{fig:smap1} and \ref{fig:smap2}  show the final images for 2.3 GHz observation in Period 1 and 2, respectively, and Figures \ref{fig:xmap1} and \ref{fig:xmap2}  show those for 8.4 GHz observation in Period 1 and 2, respectively.
Since the total observation time in the scheduled scan was almost the same for all the observations (about 9 hours), the obtained dirty beams are similar to each other.
However, dirty beams for the 2005.35 epoch were not the same because of data loss at the St. Croix station due to antenna problems. 

\section{Results}

\subsection{Proper motion of the jet}

For quantitative evaluation of the source structure and proper motion, we applied model fitting to several components for the visibility data using the MODELFIT program in DIFMAP. 
The core could be reasonably fitted with a single Gaussian component. 
On the other hand, since the jet did not show a discrete knot structure clearly in our images, Gaussian fitting of the jet region led to a single, largely extended component over the whole jet. 
In order to obtain the distribution of the individual knot positions in the jet, we applied a point source fitting. 
The procedure for model fitting of the core and jet components is shown as below. 
First we carried out the Gaussian fitting of the core. 
In the residual map subtracted the fitted components, we found the local brightest peak in the jet region and measured its position and flux as an initial model. 
Then we carried out the Gaussian fitting of the core and the point-source fitting of the jet components simultaneously. 
This process was repeated until the flux of the local peak became less than 3 $\sigma$ level. 
The $\chi^2_\nu$ values of the fitting results are calculated by using errors derived from the scatter of the visibility amplitude data within solution interval which is typically 30 seconds. 

We successfully fitted 4 or 5 components in the 2.3 GHz images and 2, 3, or 4 components in the 8.4 GHz images.

The errors for model fitting were estimated from the value that increases $\chi^2$ by 1 from the minimum value. 
Simple fitting error analysis of the flux density leads to very small errors (typically 1 mJy). 
We added flux errors to 10 $\%$ of the measured integrated flux density by assuming the typical accuracy of the absolute amplitude  calibration of the VLBA \citep{cai07}. 

The models for 2.3 GHz and 8.4 GHz are given in Tables 2 and 3, respectively. The fitted position of knots is denoted by a cross (+) and that of the core is denoted by an ellipse in the images.  
Component names in this paper are defined as follows:  in the 2.3 GHz images, components A, B, C, D, and E, from downstream of the jet to the core F$_{\rm S}$, 
and in the 8.4 GHz images, the most inner component E at 2.3 GHz is resolved into E1, E2, E3, E4, and E5, from downstream of the jet to the core F$_{\rm X}$.  

Knot identification at 2.3 GHz between Period 1 and 2 is very clear, because no systematic motion of the knots can be found over our observation period. 
Time evolution of the position shift of each component appears to be random rather than systematic, which implies that there is no evidence for proper motion in the 2.3 GHz images with 3 mas resolution. 
In contrast, knot identification at 8.4 GHz is not necessarily clear, because components E1, E2, E3 and E4 appear to show expanding motion toward downstream of the jet. 
Thus, in this subsection, we first measure proper motion at 8.4 GHz in each Period separately. We then measure the combined proper motion between Period 1 and 2 in the next subsection, 
based on the results from the proper motion in each Period.
In Tables 2 and 3, we have already referred to the identified names between Period 1 and 2, which are anticipated from the result presented in \S 3.2. 

To analyze the jet proper motion for each component, we carried out weighted least-squares fitting to the jet components detected at 8.4 GHz. 
Assuming that the core position is stationary, and that the jet moves along a straight line at an equal speed, 
we derived the fits to the R.A. direction (X) and Dec. direction (Y) separately. We then calculated proper motion using  
$u=\sqrt{u_{\rm x}^2+ u_{\rm y}^2}$, and position angle (PA) using $\phi$=tan$^{-1}$($u_{\rm x}$/$u_{\rm y}$), where $u_{\rm x}$ and $u_{\rm y}$ is the jet speed toward the X and Y directions, respectively. 
The results of our fits are shown in Table 4 and in Figures 5 and 6 for Period 1 and Period 2, respectively. 

We estimated the proper motion of E3, E2, and E1 to be 0.21$\pm$0.01, 0.38$\pm$0.02, and 0.70$\pm$0.07 mas year$^{-1}$, which corresponds to $\beta_{\rm app}$= 0.30$\pm$0.02, 0.53$\pm$0.03, and 0.96$\pm$0.09, at  0.7, 1.5, and 2.5 mas (0.3, 0.7, and 1.1 pc) from the core during Period 1, respectively,
We estimated that of E4 and E3 to be 0.23$\pm$0.03 and 0.31$\pm$0.02 mas year$^{-1}$, which corresponds $\beta_{\rm app}$= 0.32$\pm$0.04 and 0.43$\pm$0.03 at 0.7 and 1.5 mas (0.3 and 0.7 pc) from the core during Period 2, respectively. 
The PA of the proper motion is different from that of the averaged jet direction at 8.4 GHz (57 degrees, see \S 3.4)  by approximately 30 degrees for E2 and 20 degrees for E3,
implying possible curved trajectory of these knots.  

Dynamic range of our images at 8.4 GHz is not so high, typically 150. This would cause a  dominant source of systematic position error in model fitting near the edge of the jet. 
In fact, the data points at the 2002.14 epoch and the 2004.81 epoch show a large deviation from the fitted line for the outer jet components  (see Figures 5 and 6).

\subsection{Combined proper motion between Period 1 and Period 2}

In order to obtain the proper motion at 8.4 GHz over Period 1 and 2, a duration of more than 4 years, we carried out knot identification between Periods 1 and 2.
Figure 7 shows the combined proper motion of component E3.
The motion $u_{\rm x}$ of component E3 in each Period is in good agreement with each other, which demonstrates that a constant speed is kept in the jet between 0.5 and 1.6 mas (0.2 and 0.7 pc).
However,   the motion  $u_{\rm y}$ of component E3 appears to be demonstrate acceleration, suggesting a curved trajectory. 


Although tracing of the jet trajectory is very interesting for the physics of jet propagation, we do not address it further in this paper because our observations did not cover over the whole period of the movement of E3. 
We simply fitted E3 over our whole observation period using a straight line as a first-order approximation, and obtained  $u_{\rm x}=0.245\pm0.002$ mas year$^{-1}$, $u_{\rm y}=0.139\pm0.002$ mas year$^{-1}$, $u=0.282\pm0.002$ mas year$^{-1}$, $\beta_{\rm app}=0.390\pm0.003$, and $\phi=60.5\pm0.6$ degrees.

\subsection{Calibration of the proper motion using a model of orbital motion of the core}

The orbital motion of the core of 3C 66B with the orbital period of 1.05 $\pm$ 0.03 years has been shown by the phase-referencing VLBI observations in Period 1 \citep{sudou03}.
This core shift could  cause apparent position shifts of the jet components systematically, because we assumed that the core position is stationary in measuring the proper motion of the jet. 
This effect is expected to not be negligible if the position error of the knots is smaller than the major axis of the orbital motion, which was estimated to be 0.080 mas at 8.4 GHz.

Assuming that the core position in our images moves in the same way as the core orbital motion, we are able to calibrate for the effect of the core motion and to expect to obtain a more accurate measurement of the proper motion of the jet.  
The core orbital motion is  modeled on the basis of the presence of a binary black hole with a circular orbit in the core of 3C 66B, a method which was applied in \citet{sudou03}.
The obtained calibration table is presented in Table 5. 

We show the calibrated proper motion of the jet in Figures 8 and 9 and Table 6 in the same manner as Figures 5 and 6 and Table 4. 
Significant differences between the calibrated and non-calibrated proper motion could not be found. 
The fitting result of the most inner component, E3, gave a much better $\chi^2_\nu$ value than that of the non-calibrated one in both the X and Y directions.
Calibration toward the X direction at the 2002.45 epoch will work very effectively, because the core position at this epoch is isolated from those at the other epochs indicated in Table 5. 
Indeed, the decrease in the  $\chi^2_\nu$ value of E3 is mainly caused by the position shift at the 2002.45 epoch. 
The other components exhibited  $\chi^2_\nu$ values that were unchanged or worse in either or both directions.  
It is natural that the calibration of  the outer jet component, E1, with a position error larger than the core orbital diameter (0.08 mas) would be inefficient.
Calibration for components E3 and E4 in Period 2 also appear to be inefficient, even though their position errors are smaller than the orbital radius.  
Originally, their non-calibrated fit of $u_{\rm x}$ showed the $\chi^2_\nu$ value much larger than unity (58.4 and 45.0, respectively), indicating their position shift contains some unexpected error. 
This is probably why calibration is not effective for these components.

\subsection{Structure of the jet}

The images at 2.3 GHz show the straight and smooth jet structure with several knots and the unresolved core in 3C 66B at scales within a distance of 20 mas (9 pc) from the nucleus (Figures 1 and 2). 
The jet extends toward the north-east direction with the PA of 55 $\pm$ 4 degrees, which is estimated from the distribution of the knots. 
Weak wiggling patterns can be found, in particular at epoch 2001.20, 2001.48, 2002.14, and 2002.45.
These parsec-scale characteristics agree with results in the VLBI images previously obtained at 1.6 GHz and 5.0 GHz \citep{fraix97, xu00, giovannini01}. 

The images at 8.4 GHz show the sub-parsec scale jet structure at scales within a distance of 3 mas (1.3 pc) from the nucleus (Figures 3 and 4).
At this frequency, the jet becomes relatively weaker and smoother than at 2.3 GHz. 
The averaged PA of the jet is estimated to be 57 $\pm$ 4 degrees.   
The difference of the PA between 2.3 GHz and 8.4 GHz is only a few degree, indicating a well-aligned jet structure from sub-parsec to parsec scales.  
A weak bending jet pattern toward the east direction at a distance of 2 mas from the core is seen in 8.4 GHz images, in particular at epoch 2001.20, 2001.48, 2002.14, and 2005.69 (Figures 3 and 4).

We were unable to find the counterjet that was detected in the 5.0 GHz image by \citet{giovannini01}. 
Assuming that the reported jet/counterjet ratio of 10 can be applied to our images, we expect the counterjet intensity to be 0.5 mJy which is smaller than the 3-$\sigma$ noise limit of our images at both frequencies.

\subsection{Flux variation of the jet and the core}

Time evolution of the flux density of the jet components are shown in Figure 10. 
Note that the sudden flux dip at the 2005.35 epoch at both frequencies may be affected by the lower resolution due to the lack of longer baseline data related to the St. Croix station. 
All the components seem to stay at an unchanged flux level within measurement errors.
By assuming the jet flux densities are constant, we can obtain an absolute amplitude error of 15 \%. This is consistent with the typical error that we assumed in \S 2.

We also found that the flux density of the core increased by about 50 $\%$ at 8.4 GHz (see Figure 11).
This flux curve of the core is uncorrelated with those in the jet components.
Recently \citet{iguchi10} revealed the 93-day variation of the mm-wave flux density of the core region, and interpreted it as periodic Doppler modulation due to the orbital motion of a sub-parsec scale binary black hole. 
This variation is expected to be hard to be detected at cm-waves, because Syncrotron self-absorption would be dominated in the nucleus at lower radio frequencies and dilute the variation amplitude. 
Unfortunately the time sampling of our observations is too sparse and uneven (10 to 100 days), this short-term variation cannot be discussed in this paper. 

\subsection {Structural change of the core}

Figure 11 shows the changes of the integrated flux density, major axis, and PA of the fitted parameters of the core. 
At 8.4 GHz, we found that the major axis increased with time from 0.28 to 0.38 mas in Period 1, and that after reaching its largest value of 0.42 mas, decreased to 0.32 mas at the last epoch in Period 2.  
The PA changed from 80 to 50 degrees in Period 1. A possible increase of the flux density and major axis can be also seen at 2.3 GHz. 
This extension of the major axis and flux enhancement of the core can be explained qualitatively by a new component emerging that propagates in the core region.  

The axial ratio of the core at both frequencies equals to zero at almost all epochs, indicating that the core cannot be resolved toward perpendicular to the jet axis on the sub-parsec scale. 

\section{Discussion}

We plotted the apparent speed of the jet as a function of the distance from the core in Figure 12a. 3C 66B shows that the apparent jet speed increases with a distance from the core within 3 mas (1.3 pc).
This result agrees with the fact that the sub-parsec scale counterjet was found within a distance of 2 mas from the core \citep{giovannini01}.  
Since the measured proper motion in 3C 66B is considered to reflect the bulk speed of the jet within observational error \citep{iguchi10}, the apparent jet speed can be written as

\begin{equation}
 \beta_{\rm app}=\frac{\beta \sin i}{1-\beta \cos i},
\end{equation}
where $\beta$ is the jet outflow speed (=$v/c$) and $i$ is the viewing angle of the jet \citep{ginzburg69}. 
This equation indicates that increases in the apparent speed are caused by either jet acceleration or a change in the viewing angle. 
Figure 12b shows the constraints of the jet speed and the viewing angle according to Equation (1). \citet{iguchi10} revealed that $\beta=0.77$ and $i=5.3$ degrees using
the mm-wave periodic variation together with the proper motion of E3 in Period 1. 

Assuming that the jet accelerates with a constant viewing angle of 5.3 degrees,  the jet speed accelerates from 0.77 $c$ to 0.92 $c$, corresponding to the bulk Lorentz factor $\Gamma$ from 1.60 to 2.57. 
Since the relativistic energy can be written as $E=\Gamma mv^2$, the kinetic energy gain by this acceleration corresponds to 1 GeV for normal plasma and 0.6 MeV for 
pair plasma. 
This energy gain may relate to a possible $\gamma$-ray emission from 3C66B. Interestingly a very high energy (VHE) $\gamma$-ray emission above 150 GeV was detected in the region that includes both 3C 66B and a blazar 3C 66A by the MAGIC telescope \citep{aliu09}. 
\citet{tavecchino09} discussed a possible VHE $\gamma$-ray mechanism from 3C 66B in an analogy of blazar emission models. 
If indeed 3C 66B is a VHE source, the VHE $\gamma$-ray emission might be related to the sub-parsec scale acceleration of the jet.
 However, we cannot discuss any more the detailed mechanism of the $\gamma$-ray emission from 3C 66B, because physical conditions in the nuclear region are unclear from very limited SED data, e.g., a thermal protons density for proton-proton interaction or high-energy electrons density for the inverse-Compton mechanism.

Assuming the other interpretation, that is, a change in the viewing angle with a constant jet velocity of 0.77 $c$, the angle varies from 5.3 to 20.8 degrees. 
This interpretation is supported from the presence of the weak apparent bend in the jet at a distance of 2 mas from the core described in \S 3.4. 
However, there are some sources which show an apparent increase in the jet speed with a distance from pc to 100-pc scales without any bends in the jet, e.g., Cen A \citep{tingay98, hardcastle03}. 
This may indicate that the relativistic jet model expressed in equation (1) seems to be too simple to explain general properties of the apparent jet acceleration. 
We may need to consider more sophisticated models, for example, changes of emissivity profile with the distance from the core in the stratified (two-zone) jet with the spine and layer structure. 
In this case, it is expected that the spine jet is highly relativistic and that the layer jet is mildly relativistic. 
If the two-zone jet model is correct, the speed of each component would be discontinuously changed at the boundary between the region where emission from the layer jet is dominated and the region where emission from the spine jet is dominated. 
In case of 3C66B, the boundary is expected to be the distance of around 2 mas from the core. 
On the other hand, if the single-zone jet model which we assumed in equation (1) is correct, the speed of each component would be increased gradually with the distance from the core. 
For this purpose, a longer-span observation of several years is needed to trace the trajectory of the individual component in the parsec-scale jet of 3C 66B.

\section{Conclusion}

We have presented VLBI images at 2.3 GHz and 8.4 GHz at 10 epochs over 4 years. 
Subluminal motion of the jet ranging from 0.30 $c$ to 0.96 $c$ was detected within a distance of 3 mas (1.3 pc) from the core from the 8.4 GHz images. 
We demonstrated the calibration of the core position using the result of phase-referencing VLBI observations, which could improve the accuracy of measuring subluminal proper motion of a sub-parsec scale jet in nearby radio galaxies. 
Using a model of the core motion due to an orbital motion of a binary black hole found in 3C 66B, 
we calibrated the proper motion of the jet, and found no significant difference between the calibrated and non-calibrated proper motions.
Our observational results indicate that the proper motion at sub-parsec scale is likely to be increased by changes in the viewing angle, for instance, the angle varies from 5 to 21 degrees in case of a constant jet speed of 0.77 $c$. 
However, the possibility of acceleration of the jet outflow speed still remains. 
Another interpretation is also suggested that changes of emissivity profile in the two-zone jet cause the changes of the proper motion. 
To reveal the kinematics and geometry of the jet in 3C 66B, further monitoring is required over a longer period of several years to trace the individual jet components from sub-parsec to parsec scales.

\acknowledgments

This research was partially supported by
the Ministry of Education, Culture, Sports,
Science and Technology (MEXT) of Japan, Grant-in-Aid for Young Scientists (B), 16740107, 2003-2005, Grant-in-Aid 
for Young Scientists (B), 17740114, 2005-2007, and Grant-in-Aid for Scientific Research (B), 21340044, 2009-2011.

Facilities: \facility{VLBA}



\begin{deluxetable}{cllccrc}
\tablecolumns{7}
\tablewidth{0pc}
\tablecaption{Observations and image parameters}
\tablehead{
\colhead{Epoch}      & \colhead{Freq.}     & \colhead{Array}   & \colhead{Int. Time} &
\colhead{Beam Size} & \colhead{Beam PA}   & \colhead{rms noise}\\

\colhead{[yr]} & \colhead{[GHz]} &  \colhead{} & \colhead{[hr]} &
\colhead{[mas $\times$ mas]} & \colhead{[deg]}   & \colhead{[mJy/beam]}\\

\colhead{(1)} & \colhead{(2)} & \colhead{(3)} & \colhead{(4)} &
\colhead{(5)} & \colhead{(6)} & \colhead{(7)} 
}
\startdata

2001.20 	&	2.3	&	BFHKLMNOPS	&	0.7 	&	4.0 $\times$ 2.5	&	$-$1.7 	&	0.5 	\\
2001.48 	&	2.3	&	BFHKLMNOPS	&	0.3 	&	4.0 $\times$ 2.6	&	9.4 	&	0.6 	\\
2001.86 	&	2.3	&	BFHKLMNOPS	&	0.4 	&	3.7 $\times$ 2.8	&	$-$9.7 	&	0.6 	\\
2002.11 	&	2.3	&	BFHKLMNOPS	&	0.4 	&	4.0 $\times$ 2.7	&	$-$7.7 	&	0.4 	\\
2002.14 	&	2.3	&	BFHKLMNOPS	&	0.4 	&	4.0 $\times$ 2.7	&	$-$5.2 	&	0.6 	\\
2002.45 	&	2.3	&	BFHKLMNOPS	&	0.4 	&	4.0 $\times$ 2.7	&	$-$5.0 	&	0.5 	\\
2004.80 	&	2.3	&	BFHKLMNOPS	&	0.4 	&	3.9 $\times$ 2.4	&	6.8 	&	0.7 	\\
2005.05 	&	2.3	&	BFHKLMNOPS	&	0.5 	&	3.8 $\times$ 2.4	&	9.2 	&	0.7 	\\
2005.35 	&	2.3	&	BFHKLMNOPS$^\dag$	&	0.4 	&	6.4 $\times$ 3.5	&	$-$6.5 	&	0.7 	\\
2005.54 	&	2.3	&	BFHKLMNOPS	&	0.5 	&	3.8 $\times$ 2.3	&	3.2 	&	0.7 	\\
	\hline												
2001.20 	&	8.4	&	BFHKLMNOPS	&	0.6 	&	1.0 $\times$ 0.6	&	1.4 	&	0.8 	\\
2001.48 	&	8.4$^*$	&	BFHKLMNOPS	&	0.3 	&	\nodata	&	\nodata	&	\nodata	\\
2001.86 	&	8.4	&	BFHKLMNOPS	&	0.4 	&	1.1 $\times$ 0.7	&	$-$8.6 	&	0.8 	\\
2002.11 	&	8.4	&	BFHKLMNOPS	&	0.4 	&	1.0 $\times$ 0.7	&	$-$7.0 	&	0.9 	\\
2002.14 	&	8.4	&	BFHKLMNOPS	&	0.4 	&	1.0 $\times$ 0.7	&	$-$8.5 	&	0.6 	\\
2002.45 	&	8.4	&	BFHKLMNOPS	&	0.4 	&	1.1 $\times$ 0.7	&	$-$3.3 	&	0.7 	\\
2004.80 	&	8.4	&	BFHKLMNOPS	&	0.5 	&	1.1 $\times$ 0.6	&	4.8 	&	1.2 	\\
2005.05 	&	8.4	&	BFHKLMNOPS	&	0.4 	&	1.0 $\times$ 0.6	&	$-$2.0 	&	0.7 	\\
2005.35 	&	8.4	&	BFHKLMNOPS$^\dag$	&	0.4 	&	1.4 $\times$ 0.6	&	$-$21.0 	&	1.0 	\\
2005.54 	&	8.4	&	BFHKLMNOPS	&	0.5 	&	0.9 $\times$ 0.6	&	$-$3.0 	&	0.8 	\\
2005.69 	&	8.4	&	BFHKLMNOPSE	&	1.7 	&	0.8 $\times$ 0.6	&	18.4 	&	0.4 	\\

\enddata
\tablecomments{(1) Observing Epoch, (2) Observing Frequency, (3) Array configuration; B=Brewster, F=Fort Davis, H=Hancock, K=Kitt Peak,  
L=Los Alamos, M=Mauna Kea,  N=North Liberty,  O=Owens Valley,  P=Pie Town, S=St. Croix, E=Effelsberg, the diameter of the antennas is 
100 m for Effelsberg and 25 m for the VLBA antennas, (4) Integration time, (5) Major axis and minor axis of the CLEAN beam, 
(6) Position angle of the CLEAN beam, (7) rms noise level of image.\\
$^*$ We removed this data at 8.4 GHz from imaging analysis because of low quality of the data (see \S 2).\\
$^\dag$ The data of the St. Croix station was completely removed because of instrumental problems. 
 This makes the beam size about 1.5 times larger than that of the full VLBA configuration.
}

\end{deluxetable}


\begin{deluxetable}{clrrrrrr}
\rotate

\tablecolumns{8}
\tablewidth{0pc}
\tablecaption{Positions and flux densities of the knots at 2.3 GHz}
\tablehead{
\colhead{Epoch} & \colhead{Comp.}   & \colhead{$F$}    & \colhead{$\Delta{x}$} &
\colhead{$\Delta{y}$}    & \colhead{Major axis}   & \colhead{Axial}    & \colhead{PA}\\
\colhead{} & \colhead{}   & \colhead{[Jy]}    & \colhead{[mas]} &
\colhead{[mas]}    & \colhead{[mas]}   & \colhead{ratio}    & \colhead{[deg]}\\
\colhead{(1)} & \colhead{(2)} & \colhead{(3)} & \colhead{(4)} &
\colhead{(5)} & \colhead{(6)} & \colhead{(7)} & \colhead{(8)} 
}
\startdata
{  Period 1}			&		&				&				&				&				&				&				\\
2001.20 			&	F$_{\rm S}$	&	0.1012 	$\pm$	0.0102 	&	0.00 	$\pm$	0.00 	&	0.00 	$\pm$	0.00 	&	1.78 	$\pm$	0.02 	&	0.00 	$\pm$	0.00 	&	59.5 	$\pm$	0.8 	\\
($\chi^2_\nu$=	2.2 	)	&	E	&	0.0110 	$\pm$	0.0014 	&	3.99 	$\pm$	0.03 	&	1.82 	$\pm$	0.04 	&	\nodata			&	\nodata			&	\nodata			\\
			&	D	&	0.0048 	$\pm$	0.0009 	&	5.24 	$\pm$	0.08 	&	3.84 	$\pm$	0.07 	&	\nodata			&	\nodata			&	\nodata			\\
			&	C	&	0.0037 	$\pm$	0.0009 	&	8.06 	$\pm$	0.11 	&	5.80 	$\pm$	0.11 	&	\nodata			&	\nodata			&	\nodata			\\
			&	B	&	0.0048 	$\pm$	0.0009 	&	12.05 	$\pm$	0.09 	&	9.71 	$\pm$	0.10 	&	\nodata			&	\nodata			&	\nodata			\\
\hline																													
2001.48 			&	F$_{\rm S}$	&	0.1072 	$\pm$	0.0108 	&	0.00 	$\pm$	0.00 	&	0.00 	$\pm$	0.00 	&	1.23 	$\pm$	0.02 	&	0.00 	$\pm$	0.00 	&	67.2 	$\pm$	2.0 	\\
($\chi^2_\nu$=	1.8 	)	&	E	&	0.0118 	$\pm$	0.0015 	&	2.95 	$\pm$	0.04 	&	2.08 	$\pm$	0.04 	&	\nodata			&	\nodata			&	\nodata			\\
			&	D	&	0.0044 	$\pm$	0.0011 	&	5.72 	$\pm$	0.11 	&	3.44 	$\pm$	0.11 	&	\nodata			&	\nodata			&	\nodata			\\
			&	C	&	0.0038 	$\pm$	0.0010 	&	9.08 	$\pm$	0.14 	&	6.88 	$\pm$	0.14 	&	\nodata			&	\nodata			&	\nodata			\\
			&	B	&	0.0047 	$\pm$	0.0011 	&	11.97 	$\pm$	0.12 	&	9.21 	$\pm$	0.12 	&	\nodata			&	\nodata			&	\nodata			\\
			&	A	&	0.0034 	$\pm$	0.0010 	&	16.01 	$\pm$	0.14 	&	13.50 	$\pm$	0.18 	&	\nodata			&	\nodata			&	\nodata			\\
\hline																													
2001.86 			&	F$_{\rm S}$	&	0.1027 	$\pm$	0.0103 	&	0.00 	$\pm$	0.00 	&	0.04 	$\pm$	0.04 	&	1.53 	$\pm$	0.02 	&	0.00 	$\pm$	0.00 	&	58.3 	$\pm$	1.4 	\\
($\chi^2_\nu$=	2.3 	)	&	E	&	0.0135 	$\pm$	0.0015 	&	2.88 	$\pm$	0.02 	&	1.90 	$\pm$	0.02 	&	\nodata			&	\nodata			&	\nodata			\\
			&	D	&	0.0057 	$\pm$	0.0008 	&	5.38 	$\pm$	0.05 	&	4.04 	$\pm$	0.05 	&	\nodata			&	\nodata			&	\nodata			\\
			&	C	&	0.0057 	$\pm$	0.0008 	&	9.37 	$\pm$	0.06 	&	6.66 	$\pm$	0.06 	&	\nodata			&	\nodata			&	\nodata			\\
			&	B	&	0.0059 	$\pm$	0.0008 	&	12.46 	$\pm$	0.06 	&	10.49 	$\pm$	0.06 	&	\nodata			&	\nodata			&	\nodata			\\
\hline																													
2002.11 			&	F$_{\rm S}$	&	0.1211 	$\pm$	0.0121 	&	0.00 	$\pm$	0.00 	&	0.00 	$\pm$	0.00 	&	1.48 	$\pm$	0.02 	&	0.00 	$\pm$	0.00 	&	64.3 	$\pm$	1.1 	\\
($\chi^2_\nu$=	2.1 	)	&	E	&	0.0131 	$\pm$	0.0016 	&	2.57 	$\pm$	0.03 	&	1.58 	$\pm$	0.03 	&	\nodata			&	\nodata			&	\nodata			\\
			&	D	&	0.0092 	$\pm$	0.0013 	&	4.53 	$\pm$	0.06 	&	3.33 	$\pm$	0.06 	&	\nodata			&	\nodata			&	\nodata			\\
			&	C	&	0.0064 	$\pm$	0.0011 	&	8.77 	$\pm$	0.17 	&	6.30 	$\pm$	0.09 	&	\nodata			&	\nodata			&	\nodata			\\
			&	B	&	0.0065 	$\pm$	0.0011 	&	12.35 	$\pm$	0.08 	&	10.04 	$\pm$	0.09 	&	\nodata			&	\nodata			&	\nodata			\\
			&	A	&	0.0036 	$\pm$	0.0010 	&	14.98 	$\pm$	0.17 	&	13.53 	$\pm$	0.18 	&	\nodata			&	\nodata			&	\nodata			\\
\hline																													
2002.14 			&	F$_{\rm S}$	&	0.1229 	$\pm$	0.0123 	&	0.00 	$\pm$	0.00 	&	0.00 	$\pm$	0.00 	&	1.58 	$\pm$	0.02 	&	0.19 	$\pm$	0.19 	&	51.3 	$\pm$	0.8 	\\
($\chi^2_\nu$=	2.1 	)	&	E	&	0.0132 	$\pm$	0.0017 	&	2.59 	$\pm$	0.03 	&	1.66 	$\pm$	0.03 	&	\nodata			&	\nodata			&	\nodata			\\
			&	D	&	0.0076 	$\pm$	0.0013 	&	4.39 	$\pm$	0.06 	&	3.35 	$\pm$	0.06 	&	\nodata			&	\nodata			&	\nodata			\\
			&	C	&	0.0051 	$\pm$	0.0011 	&	7.00 	$\pm$	0.17 	&	5.56 	$\pm$	0.15 	&	\nodata			&	\nodata			&	\nodata			\\
			&	B	&	0.0062 	$\pm$	0.0012 	&	12.21 	$\pm$	0.08 	&	9.98 	$\pm$	0.09 	&	\nodata			&	\nodata			&	\nodata			\\
			&	A	&	0.0028 	$\pm$	0.0010 	&	15.75 	$\pm$	0.17 	&	14.73 	$\pm$	0.19 	&	\nodata			&	\nodata			&	\nodata			\\
\hline																													
2002.45			&	F$_{\rm S}$	&	0.1137 	$\pm$	0.0114 	&	0.00 	$\pm$	0.00 	&	0.00 	$\pm$	0.00 	&	1.64 	$\pm$	0.02 	&	0.00 	$\pm$	0.00 	&	59.6 	$\pm$	1.0 	\\
($\chi^2_\nu$=	2.1 	)	&	E	&	0.0123 	$\pm$	0.0015 	&	2.85 	$\pm$	0.03 	&	2.06 	$\pm$	0.04 	&	\nodata			&	\nodata			&	\nodata			\\
			&	D	&	0.0056 	$\pm$	0.0011 	&	4.93 	$\pm$	0.08 	&	2.99 	$\pm$	0.08 	&	\nodata			&	\nodata			&	\nodata			\\
			&	C	&	0.0068 	$\pm$	0.0012 	&	8.40 	$\pm$	0.07 	&	6.12 	$\pm$	0.07 	&	\nodata			&	\nodata			&	\nodata			\\
			&	B	&	0.0063 	$\pm$	0.0011 	&	11.72 	$\pm$	0.09 	&	9.14 	$\pm$	0.09 	&	\nodata			&	\nodata			&	\nodata			\\
			&	A	&	0.0031 	$\pm$	0.0010 	&	15.08 	$\pm$	0.14 	&	13.29 	$\pm$	0.16 	&	\nodata			&	\nodata			&	\nodata			\\
\hline																													
{  Period 2}			&		&				&				&				&				&				&				\\
2004.80 			&	F$_{\rm S}$	&	0.1108 	$\pm$	0.0111 	&	0.00 	$\pm$	0.00 	&	0.00 	$\pm$	0.00 	&	1.69 	$\pm$	0.01 	&	0.00 	$\pm$	0.00 	&	64.4 	$\pm$	1.1 	\\
($\chi^2_\nu$=	2.1 	)	&	E	&	0.0121 	$\pm$	0.0015 	&	2.87 	$\pm$	0.04 	&	2.12 	$\pm$	0.04 	&	\nodata			&	\nodata			&	\nodata			\\
			&	D	&	0.0047 	$\pm$	0.0010 	&	4.99 	$\pm$	0.10 	&	2.86 	$\pm$	0.10 	&	\nodata			&	\nodata			&	\nodata			\\
			&	C	&	0.0059 	$\pm$	0.0011 	&	8.44 	$\pm$	0.08 	&	6.02 	$\pm$	0.08 	&	\nodata			&	\nodata			&	\nodata			\\
			&	B	&	0.0045 	$\pm$	0.0010 	&	11.84 	$\pm$	0.13 	&	9.42 	$\pm$	0.13 	&	\nodata			&	\nodata			&	\nodata			\\
\hline																													
2005.05 			&	F$_{\rm S}$	&	0.1011 	$\pm$	0.0101 	&	0.00 	$\pm$	0.00 	&	0.00 	$\pm$	0.00 	&	1.70 	$\pm$	0.02 	&	0.00 	$\pm$	0.00 	&	58.7 	$\pm$	0.9 	\\
($\chi^2_\nu$=	2.0 	)	&	E	&	0.0122 	$\pm$	0.0014 	&	2.71 	$\pm$	0.03 	&	1.49 	$\pm$	0.03 	&	\nodata			&	\nodata			&	\nodata			\\
			&	D	&	0.0044 	$\pm$	0.0008 	&	4.51 	$\pm$	0.08 	&	3.35 	$\pm$	0.08 	&	\nodata			&	\nodata			&	\nodata			\\
			&	C	&	0.0046 	$\pm$	0.0009 	&	7.49 	$\pm$	0.08 	&	5.21 	$\pm$	0.08 	&	\nodata			&	\nodata			&	\nodata			\\
			&	B	&	0.0053 	$\pm$	0.0009 	&	11.41 	$\pm$	0.09 	&	8.76 	$\pm$	0.08 	&	\nodata			&	\nodata			&	\nodata			\\
\hline																													
2005.35 			&	F$_{\rm S}$	&	0.1157 	$\pm$	0.0116 	&	0.00 	$\pm$	0.00 	&	0.00 	$\pm$	0.00 	&	2.36 	$\pm$	0.04 	&	0.00 	$\pm$	0.00 	&	50.9 	$\pm$	1.3 	\\
($\chi^2_\nu$=	2.0 	)	&	E	&	0.0067 	$\pm$	0.0013 	&	2.72 	$\pm$	0.09 	&	1.99 	$\pm$	0.09 	&	\nodata			&	\nodata			&	\nodata			\\
			&	D	&	0.0086 	$\pm$	0.0014 	&	4.46 	$\pm$	0.11 	&	3.00 	$\pm$	0.12 	&	\nodata			&	\nodata			&	\nodata			\\
			&	C	&	0.0057 	$\pm$	0.0012 	&	8.59 	$\pm$	0.10 	&	6.09 	$\pm$	0.11 	&	\nodata			&	\nodata			&	\nodata			\\
			&	B	&	0.0074 	$\pm$	0.0013 	&	12.68 	$\pm$	0.12 	&	9.42 	$\pm$	0.13 	&	\nodata			&	\nodata			&	\nodata			\\
\hline																													
2005.54 			&	F$_{\rm S}$	&	0.1271 	$\pm$	0.0127 	&	0.00 	$\pm$	0.00 	&	0.00 	$\pm$	0.00 	&	1.69 	$\pm$	0.02 	&	0.00 	$\pm$	0.00 	&	53.9 	$\pm$	0.9 	\\
($\chi^2_\nu$=	2.3 	)	&	E	&	0.0162 	$\pm$	0.0019 	&	2.92 	$\pm$	0.03 	&	1.63 	$\pm$	0.03 	&	\nodata			&	\nodata			&	\nodata			\\
			&	D	&	0.0059 	$\pm$	0.0011 	&	5.10 	$\pm$	0.08	&	3.10 	$\pm$	0.08 	&	\nodata			&	\nodata			&	\nodata			\\
			&	C	&	0.0060 	$\pm$	0.0011 	&	8.71 	$\pm$	0.09 	&	5.95 	$\pm$	0.09 	&	\nodata			&	\nodata			&	\nodata			\\
			&	B	&	0.0060 	$\pm$	0.0011 	&	12.00 	$\pm$	0.10 	&	9.03 	$\pm$	0.10 	&	\nodata			&	\nodata			&	\nodata			\\\enddata

\tablecomments{(1) Observing Epoch, (2) Component names defined as follows;  components A, B, C, D, and E, from downstream of the jet to the core F$_{\rm S}$, 
 (3) Flux density, (4) R.A. offset, (5) Dec. offset, (6) Major axis of Gaussian components, (7) Axial ratio of Gaussian components, (8) Position angle of Gaussian components.}

\end{deluxetable}


\begin{deluxetable}{clrrrrrr}
\rotate
\tablecolumns{8}
\tablewidth{0pc}
\tablecaption{Positions and flux densities of the knots at 8.4 GHz}
\tablehead{
\colhead{Epoch} & \colhead{Comp.}   & \colhead{$F$}    & \colhead{$\Delta{x}$} &
\colhead{$\Delta{y}$}    & \colhead{Major axis}   & \colhead{Axial}    & \colhead{PA}\\
\colhead{} & \colhead{}   & \colhead{[Jy]}    & \colhead{[mas]} &
\colhead{[mas]}    & \colhead{[mas]}   & \colhead{ratio}    & \colhead{[deg]}\\
\colhead{(1)} & \colhead{(2)} & \colhead{(3)} & \colhead{(4)} &
\colhead{(5)} & \colhead{(6)} & \colhead{(7)} & \colhead{(8)} 
}
\startdata

{  Period 1}			&		&				&				&				&				&				&				\\
2001.20 			&	F$_{\rm X}$	&	0.088 	$\pm$	0.009 	&	0.00 	$\pm$	0.00 	&	0.00 	$\pm$	0.00 	&	0.28 	$\pm$	0.01 	&	0.00 	$\pm$	0.00 	&	78.0 	$\pm$	1.6 	\\
($\chi^2_\nu$=	1.0 	)	&	E3	&	0.015 	$\pm$	0.002 	&	0.50 	$\pm$	0.01 	&	0.42 	$\pm$	0.01 	&	\nodata			&	\nodata			&	\nodata			\\
			&	E2	&	0.007 	$\pm$	0.001 	&	1.10 	$\pm$	0.02 	&	0.86 	$\pm$	0.02 	&	\nodata			&	\nodata			&	\nodata			\\
			&	E1	&	0.005 	$\pm$	0.001 	&	1.85 	$\pm$	0.05 	&	1.03 	$\pm$	0.05 	&	\nodata			&	\nodata			&	\nodata			\\
\hline																													
2001.86 			&	F$_{\rm X}$	&	0.108 	$\pm$	0.011 	&	0.00 	$\pm$	0.00 	&	0.00 	$\pm$	0.00 	&	0.31 	$\pm$	0.01 	&	0.00 	$\pm$	0.00 	&	58.3 	$\pm$	1.6 	\\
($\chi^2_\nu$=	0.9 	)	&	E3	&	0.018 	$\pm$	0.002 	&	0.65 	$\pm$	0.01 	&	0.44 	$\pm$	0.01 	&	\nodata			&	\nodata			&	\nodata			\\
			&	E2	&	0.008 	$\pm$	0.001 	&	1.32 	$\pm$	0.02 	&	0.89 	$\pm$	0.02 	&	\nodata			&	\nodata			&	\nodata			\\
			&	E1	&	0.003 	$\pm$	0.001 	&	2.13 	$\pm$	0.07 	&	1.22 	$\pm$	0.07 	&	\nodata			&	\nodata			&	\nodata			\\
\hline																													
2002.11 			&	F$_{\rm X}$	&	0.108 	$\pm$	0.011 	&	0.00 	$\pm$	0.00 	&	0.00 	$\pm$	0.00 	&	0.37 	$\pm$	0.01 	&	0.19 	$\pm$	0.08 	&	63.7 	$\pm$	3.0 	\\
($\chi^2_\nu$=	1.2 	)	&	E3	&	0.017 	$\pm$	0.002 	&	0.71 	$\pm$	0.01 	&	0.41 	$\pm$	0.01 	&	\nodata			&	\nodata			&	\nodata			\\
			&	E2	&	0.008 	$\pm$	0.001 	&	1.44 	$\pm$	0.02 	&	0.84 	$\pm$	0.02 	&	\nodata			&	\nodata			&	\nodata			\\
			&	E1	&	0.003 	$\pm$	0.001 	&	2.43 	$\pm$	0.08 	&	1.45 	$\pm$	0.08 	&	\nodata			&	\nodata			&	\nodata			\\
\hline																													
2002.14 			&	F$_{\rm X}$	&	0.119 	$\pm$	0.012 	&	0.00 	$\pm$	0.00 	&	0.00 	$\pm$	0.00 	&	0.34 	$\pm$	0.01 	&	0.08 	$\pm$	0.08 	&	63.5 	$\pm$	3.1 	\\
($\chi^2_\nu$=	0.9 	)	&	E3	&	0.019 	$\pm$	0.002 	&	0.72 	$\pm$	0.01 	&	0.42 	$\pm$	0.01 	&	\nodata			&	\nodata			&	\nodata			\\
			&	E2	&	0.008 	$\pm$	0.001 	&	1.37 	$\pm$	0.02 	&	0.83 	$\pm$	0.02 	&	\nodata			&	\nodata			&	\nodata			\\
			&	E1	&	0.003 	$\pm$	0.001 	&	2.20 	$\pm$	0.09 	&	0.95 	$\pm$	0.09 	&	\nodata			&	\nodata			&	\nodata			\\
\hline																													
2002.45 			&	F$_{\rm X}$	&	0.127 	$\pm$	0.013 	&	0.00 	$\pm$	0.00 	&	0.00 	$\pm$	0.00 	&	0.38 	$\pm$	0.02 	&	0.00 	$\pm$	0.00 	&	48.7 	$\pm$	4.1 	\\
($\chi^2_\nu$=	1.0 	)	&	E3	&	0.018 	$\pm$	0.002 	&	0.71 	$\pm$	0.02 	&	0.48 	$\pm$	0.02 	&	\nodata			&	\nodata			&	\nodata			\\
			&	E2	&	0.010 	$\pm$	0.001 	&	1.61 	$\pm$	0.02 	&	0.85 	$\pm$	0.02 	&	\nodata			&	\nodata			&	\nodata			\\
			&	E1	&	0.004 	$\pm$	0.001 	&	2.75 	$\pm$	0.09 	&	1.50 	$\pm$	0.09 	&	\nodata			&	\nodata			&	\nodata			\\
\hline																													
{  Period 2}			&		&				&				&				&				&				&				\\
2004.80 			&	F$_{\rm X}$	&	0.095 	$\pm$	0.010 	&	0.00 	$\pm$	0.00 	&	0.00 	$\pm$	0.00 	&	0.39 	$\pm$	0.02 	&	0.00 	$\pm$	0.00 	&	51.6 	$\pm$	3.3 	\\
($\chi^2_\nu$=	1.9 	)	&	E4	&	0.018 	$\pm$	0.002 	&	0.78 	$\pm$	0.02 	&	0.49 	$\pm$	0.02 	&	\nodata			&	\nodata			&	\nodata			\\
			&	E3	&	0.007 	$\pm$	0.001 	&	1.58 	$\pm$	0.02 	&	0.99 	$\pm$	0.02 	&	\nodata			&	\nodata			&	\nodata			\\
			&	E2	&	0.005 	$\pm$	0.001 	&	3.04 	$\pm$	0.09 	&	1.91 	$\pm$	0.08 	&	\nodata			&	\nodata			&	\nodata			\\
\hline																													
2005.05 			&	F$_{\rm X}$	&	0.092 	$\pm$	0.009 	&	0.00 	$\pm$	0.00 	&	0.00 	$\pm$	0.00 	&	0.42 	$\pm$	0.01 	&	0.00 	$\pm$	0.00 	&	54.0 	$\pm$	3.3 	\\
($\chi^2_\nu$=	1.9 	)	&	E4	&	0.018 	$\pm$	0.002 	&	0.68 	$\pm$	0.02 	&	0.51 	$\pm$	0.02 	&	\nodata			&	\nodata			&	\nodata			\\
			&	E3	&	0.010 	$\pm$	0.001 	&	1.39 	$\pm$	0.01 	&	0.85 	$\pm$	0.01 	&	\nodata			&	\nodata			&	\nodata			\\
\hline																													
2005.35 			&	F$_{\rm X}$	&	0.091 	$\pm$	0.009 	&	0.00 	$\pm$	0.00 	&	0.00 	$\pm$	0.00 	&	0.33 	$\pm$	0.02 	&	0.00 	$\pm$	0.00 	&	46.0 	$\pm$	6.8 	\\
($\chi^2_\nu$=	2.0 	)	&	E4	&	0.012 	$\pm$	0.002 	&	0.71 	$\pm$	0.03 	&	0.59 	$\pm$	0.03 	&	\nodata			&	\nodata			&	\nodata			\\
			&	E3	&	0.007 	$\pm$	0.001 	&	1.57 	$\pm$	0.03 	&	1.07 	$\pm$	0.03 	&	\nodata			&	\nodata			&	\nodata			\\
\hline																													
2005.54 			&	F$_{\rm X}$	&	0.120 	$\pm$	0.012 	&	0.00 	$\pm$	0.00 	&	0.00 	$\pm$	0.00 	&	0.40 	$\pm$	0.01 	&	0.13 	$\pm$	0.07 	&	57.8 	$\pm$	2.7 	\\
($\chi^2_\nu$=	1.9 	)	&	E4	&	0.020 	$\pm$	0.002 	&	0.69 	$\pm$	0.02 	&	0.48 	$\pm$	0.02 	&	\nodata			&	\nodata			&	\nodata			\\
			&	E3	&	0.010 	$\pm$	0.001 	&	1.47 	$\pm$	0.02 	&	0.94 	$\pm$	0.02 	&	\nodata			&	\nodata			&	\nodata			\\
\hline																													
2005.69 			&	F$_{\rm X}$	&	0.111 	$\pm$	0.011 	&	0.00 	$\pm$	0.00 	&	0.00 	$\pm$	0.00 	&	0.32 	$\pm$	0.01 	&	0.00 	$\pm$	0.00 	&	56.2 	$\pm$	2.1 	\\
($\chi^2_\nu$=	1.2 	)	&	E5	&	0.020 	$\pm$	0.002 	&	0.53 	$\pm$	0.02 	&	0.31 	$\pm$	0.02 	&	\nodata			&	\nodata			&	\nodata			\\
			&	E4	&	0.010 	$\pm$	0.001 	&	1.00 	$\pm$	0.02 	&	0.70 	$\pm$	0.02 	&	\nodata			&	\nodata			&	\nodata			\\
			&	E3	&	0.006 	$\pm$	0.001 	&	1.66 	$\pm$	0.01 	&	1.02 	$\pm$	0.01 	&	\nodata			&	\nodata			&	\nodata			\\
\enddata

\tablecomments{Same as Table 2, but at 8.4 GHz. Component names defined as follows;  components E1, E2, E3, E4 and E5 from downstream of the jet to the core F$_{\rm X}$.  
}

\end{deluxetable}


\begin{deluxetable}{crrrrl}
\tablecolumns{6}
\tablewidth{0pc}
\tablecaption{Proper motion of the jet in 3C66B}
\tablehead{
\colhead{Comp.} & \colhead{$u_{\rm x}$}   & \colhead{$u_{\rm y}$} & 
\colhead{$u$} & \colhead{$\beta_{\rm app}$}   & \colhead{PA}   \\
\colhead{} & \colhead{[mas/yr]}    &\colhead{[mas/yr]}    &
 \colhead{[mas/yr]} & \colhead{} & \colhead{[deg]}   \\
\colhead{(1)} & \colhead{(2)} & \colhead{(3)} & \colhead{(4)} &\colhead{(5)} & \colhead{(6)} 
}
\startdata
 {  Period 1}	&				&					&				&				&				\\
E1	&	0.60 	$\pm$	0.07 	&		0.35 	$\pm$	0.08 	&		0.70 	$\pm$	0.07 	&	0.96 	$\pm$	0.09 	&	60 	$\pm$	8 	\\
E2	&	0.38 	$\pm$	0.02 	&		$-$0.02 	$\pm$	0.02 	&		0.38 	$\pm$	0.02 	&	0.53 	$\pm$	0.03 	&	94 	$\pm$	3 	\\
E3	&	0.21 	$\pm$	0.01 	&		0.01 	$\pm$	0.01 	&		0.21 	$\pm$	0.01 	&	0.30 	$\pm$	0.02 	&	88 	$\pm$	3 	\\
\hline
{  Period 2}	&				&					& 				&				&				\\																									
E3	&	0.27 	$\pm$	0.02 	&		0.16 	$\pm$	0.02 	&		0.31 	$\pm$	0.02 	&	0.43 	$\pm$	0.03 	&	59 	$\pm$	5 	\\																								
E4	&	0.17 	$\pm$	0.03 	&		0.16 	$\pm$	0.03 	&		0.23	$\pm$	0.03	&	0.32	$\pm$	0.04 	&	46	$\pm$	10	\\

\enddata

\tablecomments{(1) Component names defined as follows;  components E1, E2, E3, and E4 from downstream of the jet to the core F$_{\rm X}$,
(2) Jet proper motion toward R.A. direction, (3) Jet proper motion toward Dec. direction, (4) Jet proper motion, (5) Apparent jet velocity $\beta_{\rm app}=v_{\rm app}/c$, (6) Position angle of the proper motion}

\end{deluxetable}


\begin{deluxetable}{crr}
\tablecolumns{3}
\tablewidth{0pc}
\tablecaption{Calibration table of the core position shift using  a model of a binary black hole}
\tablehead{
\colhead{Epoch} & \colhead{$\Delta x$}   & \colhead{$\Delta y$} \\
\colhead{[year]} & \colhead{[mas]}    &\colhead{[mas]}    \\
\colhead{(1)} & \colhead{(2)} & \colhead{(3)}  
}
\startdata
 {  Period 1}	&				&					\\
2001.20 	&	$-$0.010 	&	0.010 	\\
2001.86 	&	$-$0.024 	&	$-$0.009 	\\
2002.11 	&	$-$0.040 	&	0.005 	\\
2002.14 	&	$-$0.035 	&	0.006 	\\
2002.45 	&	0.038 	&	0.006 	\\
 {  Period 2}	&				&					\\
2004.80 	&	0.027 	&	$-$0.008 	\\
2005.05 	&	$-$0.034 	&	$-$0.008 	\\
2005.35 	&	$-$0.026 	&	0.008 	\\
2005.54 	&	0.025 	&	0.009 	\\
2005.69 	&	0.045 	&	0.002 	\\

\enddata

\tablecomments{(1)Observation epoch, (2) R. A. offset from the center of the core orbit, (3) Dec. offset from the center of the core orbit. These parameters are calculated 
based on Table 1 of \citet{sudou03}. }
\end{deluxetable}


\begin{deluxetable}{crrrrr}
\tablecolumns{6}
\tablewidth{0pc}
\tablecaption{Proper motion of the jet in 3C66B calibrated by a model of core motion}
\tablehead{
\colhead{Comp.} & \colhead{$u_{\rm x}$}   & \colhead{$u_{\rm y}$} & 
\colhead{$u$} & \colhead{$\beta_{\rm app}$}   & \colhead{PA}   \\
\colhead{} & \colhead{[mas/yr]}    &\colhead{[mas/yr]}    &
 \colhead{[mas/yr]} & \colhead{} & \colhead{[deg]}   \\
\colhead{(1)} & \colhead{(2)} & \colhead{(3)} & \colhead{(4)} &\colhead{(5)} & \colhead{(6)} 
}
\startdata
 {  Period 1}	&				&					&				&				&				\\
E1	&	0.61 	$\pm$	0.07 	&		0.35 	$\pm$	0.08 	&		0.70 	$\pm$	0.07 	&	0.96 	$\pm$	0.09 	&	61 	$\pm$	8 	\\
E2	&	0.40 	$\pm$	0.02 	&		$-$0.02 	$\pm$	0.02 	&		0.40 	$\pm$	0.02 	&	0.55 	$\pm$	0.03 	&	95 	$\pm$	3 	\\
E3	&	0.20 	$\pm$	0.01 	&		0.01 	$\pm$	0.01 	&		0.20 	$\pm$	0.01 	&	0.28 	$\pm$	0.02 	&	88 	$\pm$	3 	\\
\hline
 {  Period 2}	&				&					&				&				&				\\																									
E3	&	0.34 	$\pm$	0.02 	&		0.17 	$\pm$	0.02 	&		0.39 	$\pm$	0.02 	&	0.53 	$\pm$	0.03 	&	63 	$\pm$	4 	\\																								
E4	&	0.20 	$\pm$	0.03 	&		0.18 	$\pm$	0.03 	&		0.27	$\pm$	0.03	&	0.37	$\pm$	0.04 	&	49	$\pm$	8	\\

\enddata

\tablecomments{(1) Component names defined as follows;  components E1, E2, E3, and E4 from downstream of the jet to the core F$_{\rm X}$,
(2) Jet proper motion toward R.A. direction, (3) Jet proper motion toward Dec. direction, (4) Jet proper motion, (5) Apparent jet velocity $\beta_{\rm app}=v_{\rm app}/c$, (6) Position angle of the proper motion.}

\end{deluxetable}


\begin{deluxetable}{crrrr}
\tablecolumns{5}
\tablewidth{0pc}
\tablecaption{$\chi^2_\nu$ values of the linear fitting of proper motions before and after applying the calibration for the core motion.}
\tablehead{
\colhead{Comp.} & \multicolumn{2}{c}{X-direction}   & \multicolumn{2}{c}{Y-direction}   \\
\colhead{} & \colhead{N}    &\colhead{C}    &  \colhead{N} & \colhead{C}   \\
\colhead{(1)} & \colhead{(2)} & \colhead{(3)} & \colhead{(4)} &\colhead{(5)}
 }
\startdata
 {  Period 1}	&				&					&				&				\\
E1	&	2.1	&		4.0  	&	3.5	 &	5.0 	\\
E2	&	6.4	&		16.2 	&	1.2 	&	0.7 	\\
E3	&	4.1	&		0.1	&	3.6	&	3.2	\\
\hline
 {  Period 2}	&				&					&				&				\\																									
E3	&	58.4	&		91.0 	&	30.2 	&	31.2 	\\																								
E4	&	45.0	&		62.0	&	16.7	&	15.0	\\

\enddata

\tablecomments{(1) Component name defined in Table 4, 
(2) $\chi^2_\nu$ value for R.A. direction without the correction of the core orbital motion, (3) $\chi^2_\nu$ value for R.A. direction with the correction of the core orbital motion, 
(4) $\chi^2_\nu$ value for Dec. direction without the correction of the core orbital motion, (4) $\chi^2_\nu$ value for Dec. direction with the correction of the core orbital motion.
Only the most inner component, E3, exhibited a better $\chi^2_\nu$ value in both the R.A. and Dec. directions,  
the others exhibited unchanged or worse $\chi^2_\nu$ values in either or both directions. }

\end{deluxetable}

\onecolumn

\begin{figure}
  \begin{center}
 \epsscale{0.75}
\plotone{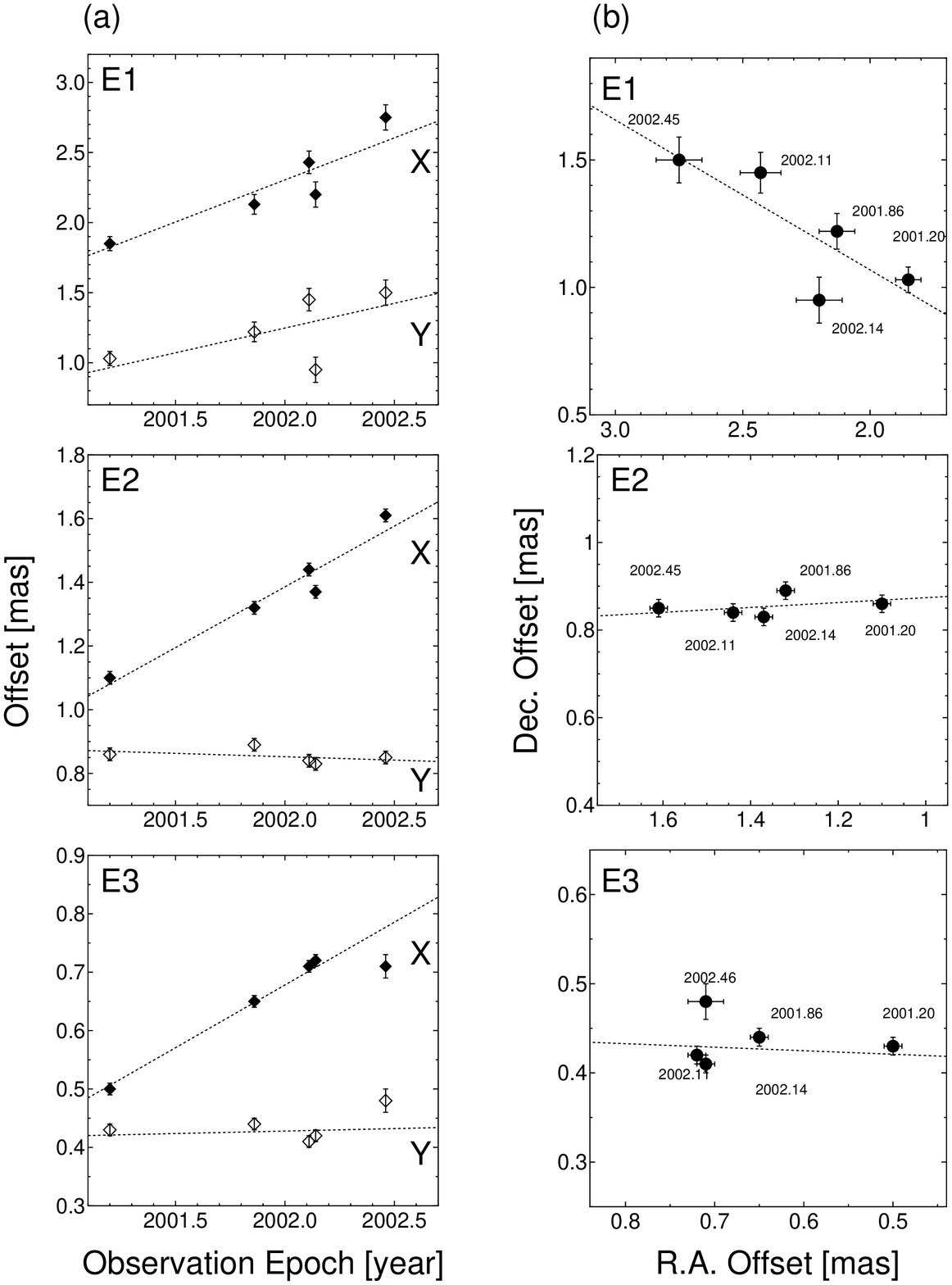} 
  \caption{Linear fitting of the relative position shift of the knots E1, E2, and E3 at 8.4 GHz in Period 1.  (a)  time evolution and (b) spatial distribution.
 A filled lozenge means the R.A. direction (X), and an open lozenge means the Dec. direction (Y). Dotted lines indicate the best fit.}
 \end{center}
\end{figure}


\begin{figure}
  \begin{center}
\epsscale{0.75}
 \plotone{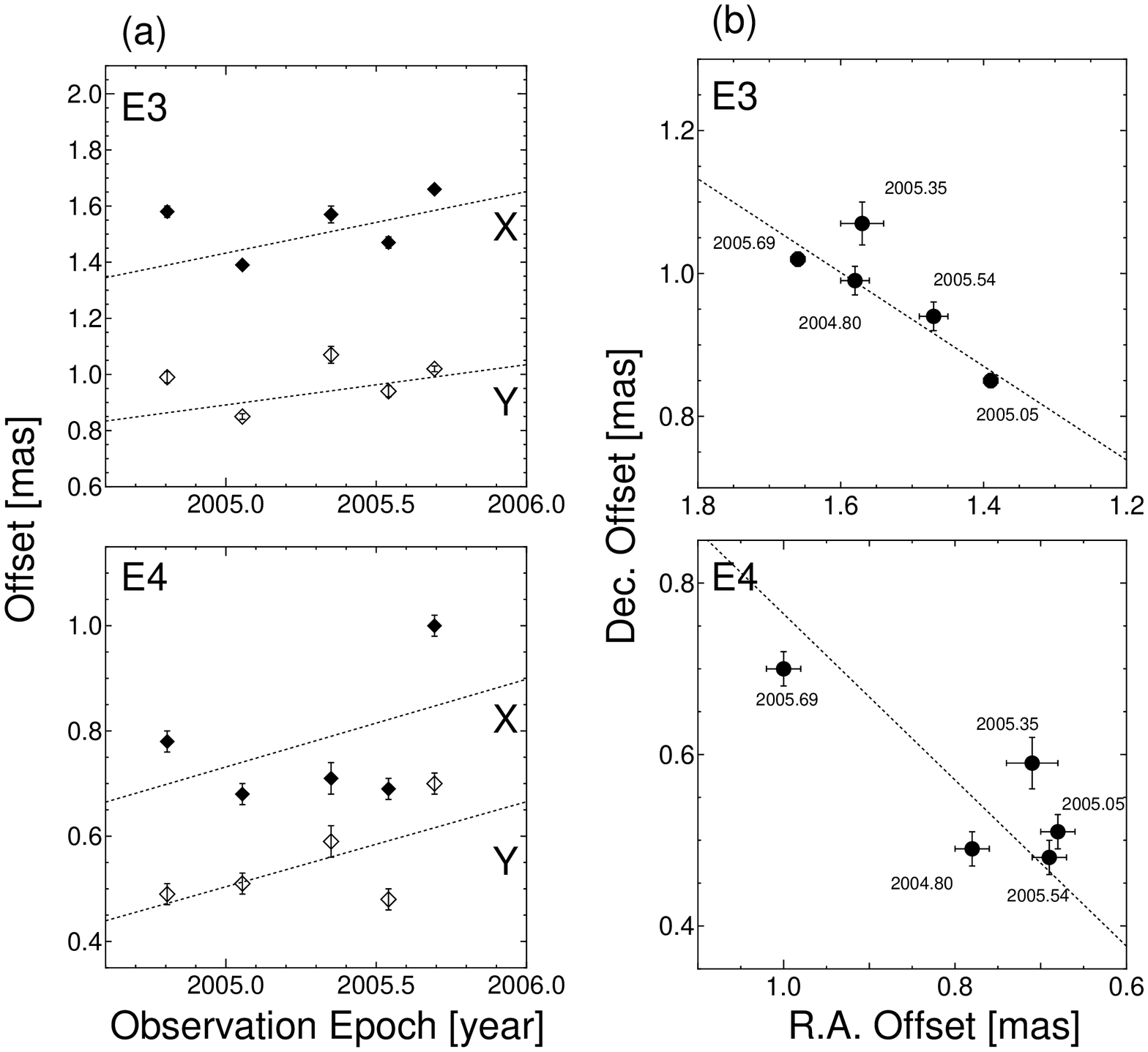} 
  \caption{ Linear fitting of the relative position shift of the knots E3 and E4  at 8.4 GHz in Period 2.  (a)  time evolution and (b) spatial distribution.
 A filled lozenge means the R.A. direction (X), and an open lozenge means the Dec. direction (Y). Dotted lines indicate the best fit.}
 \end{center}
\end{figure}


\begin{figure}
  \begin{center}
\epsscale{0.5}
 \plotone{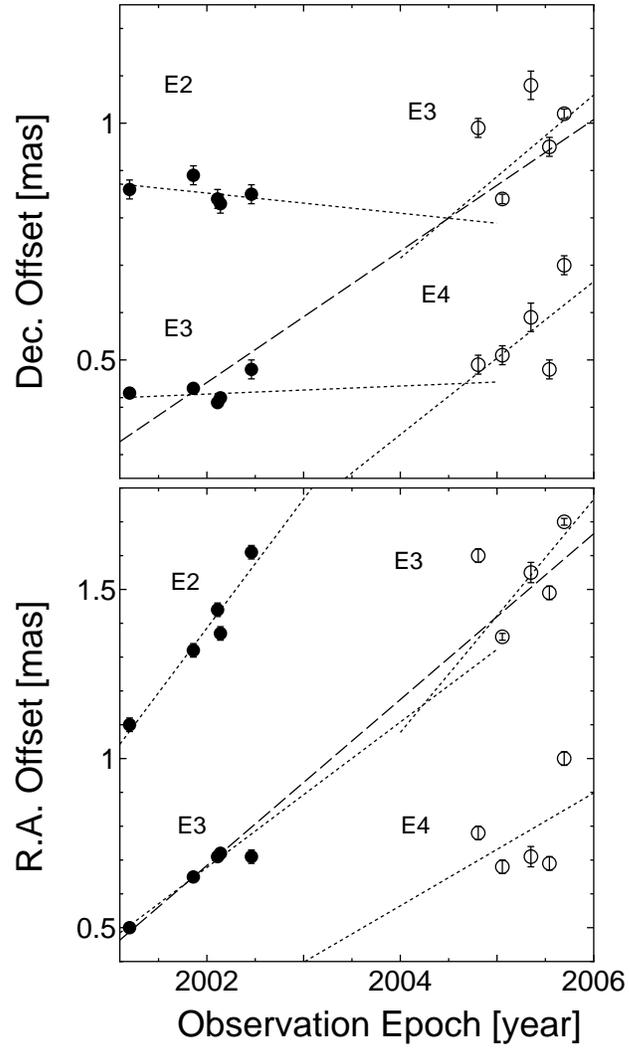}
  \caption{Combined proper motion of the component E3 between Period 1 and Period 2. Dotted lines indicate the fitting result in each Period, which was also shown in Figures 5 and 6, and the dashed line indicates the linearly fitted line of combined data of E3 between Period 1 and Period 2.  }
 \end{center}
\end{figure}

\begin{figure}
  \begin{center}
\epsscale{0.75}
\plotone{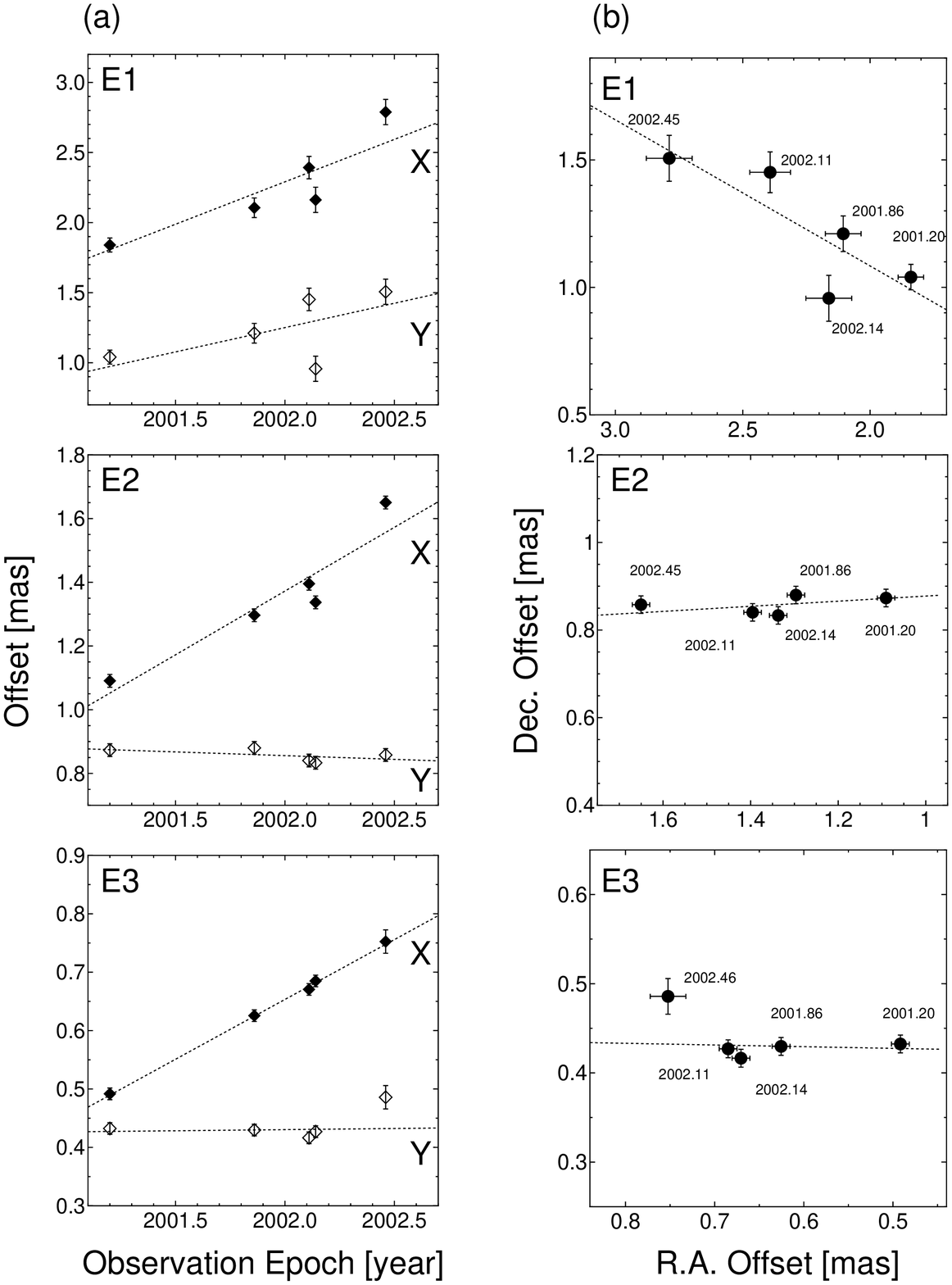} 
  \caption{Same as Figure 5, but the core position is calibrated by a model of circular orbital motion in a putative binary black hole in 3C 66B.  (a)  time evolution and (b) spatial distribution.
 A filled lozenge means the R.A. direction (X), and an open lozenge means the Dec. direction (Y). Dotted lines indicate the best fit.}
 \end{center}
\end{figure}


\begin{figure}
  \begin{center}
\epsscale{0.75}
 \plotone{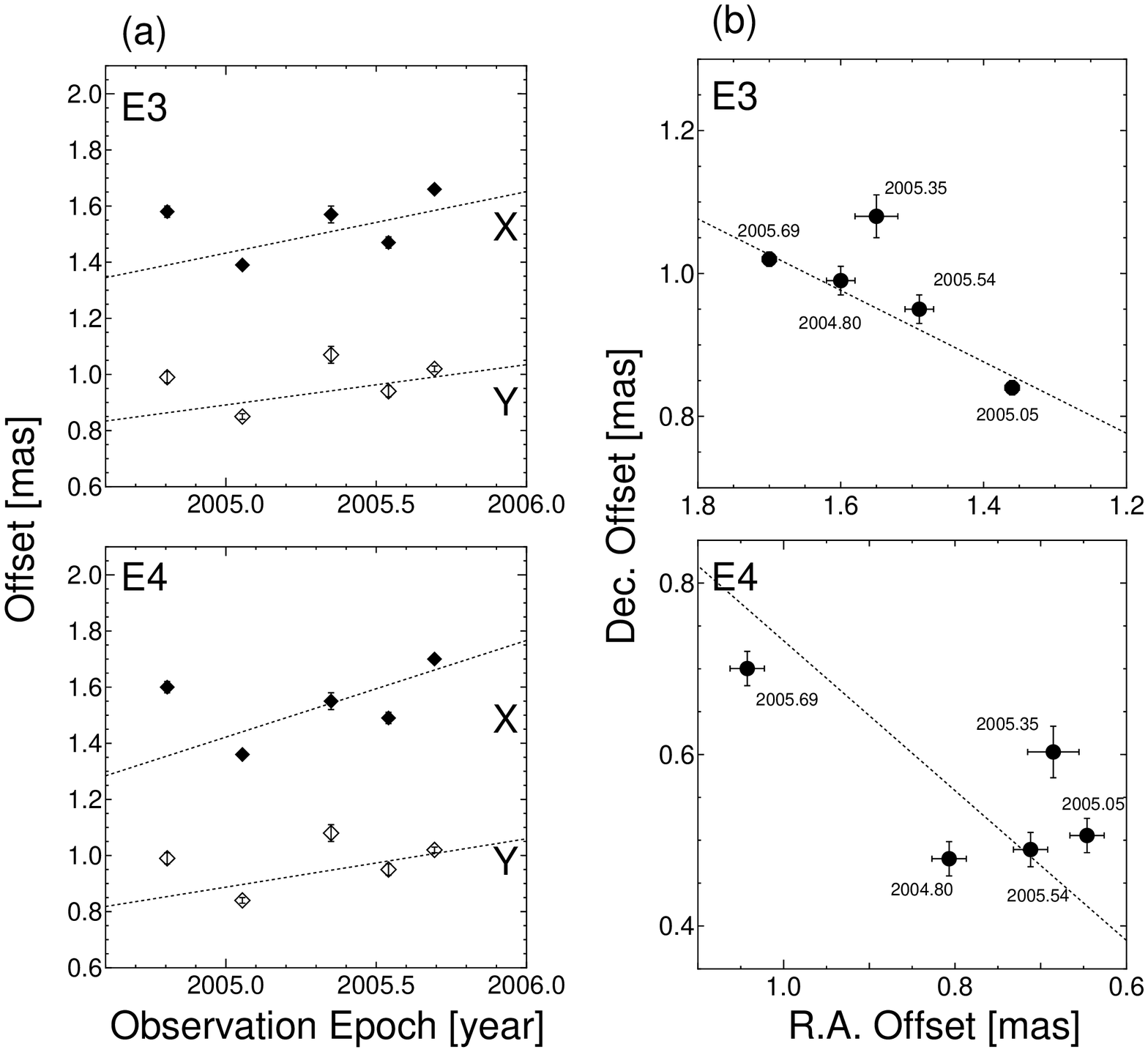} 
  \caption{Same as Figure 6, but the core position is calibrated by a model of circular orbital motion in a putative binary black hole in 3C 66B.  (a)  time evolution and (b) spatial distribution.
 A filled lozenge means the R.A. direction (X), and an open lozenge means the Dec. direction (Y). Dotted lines indicate the best fit.}
 \end{center}
\end{figure}

\begin{figure}
  \begin{center}
\epsscale{0.75}
 \plotone{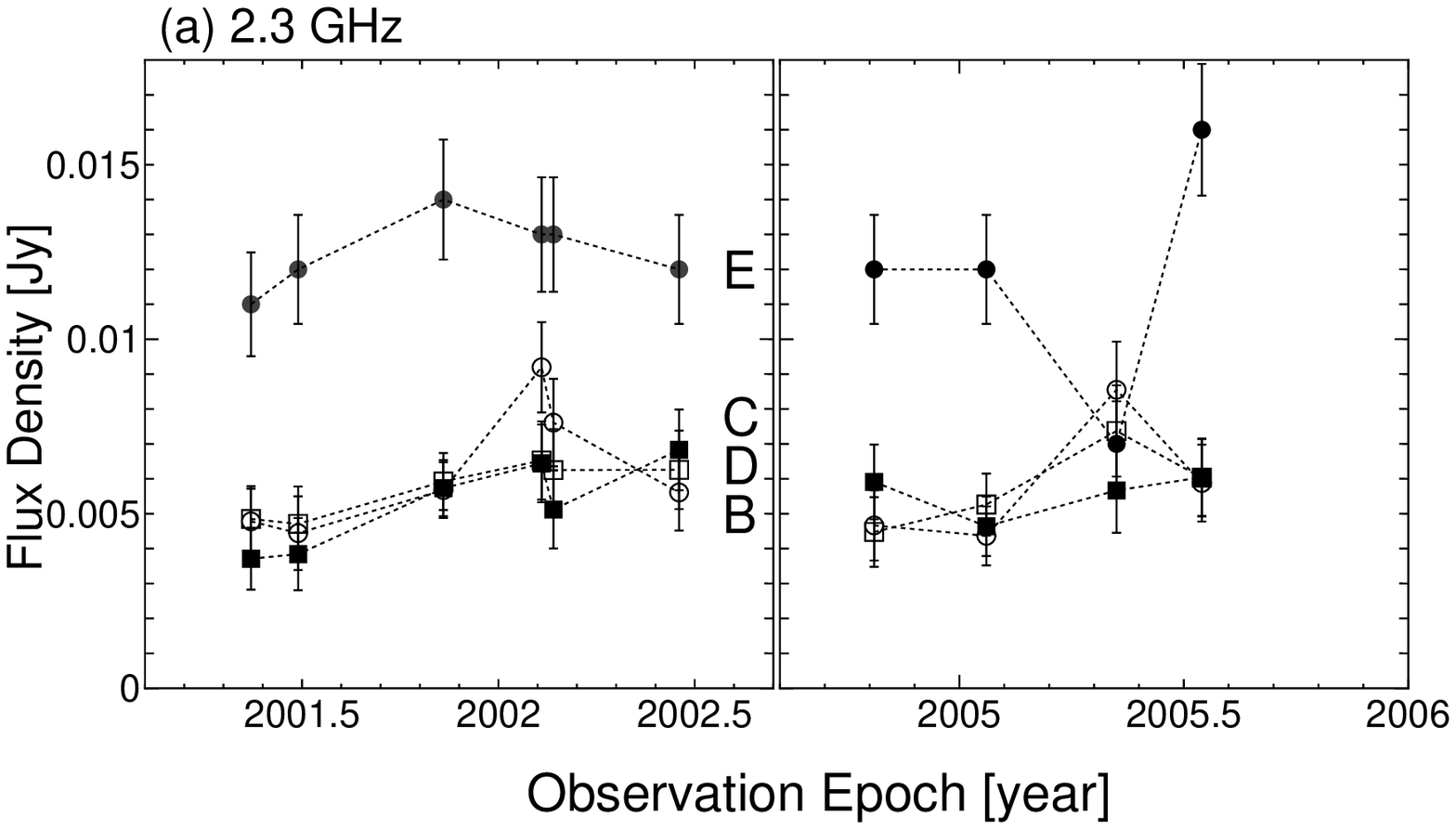}
 \plotone{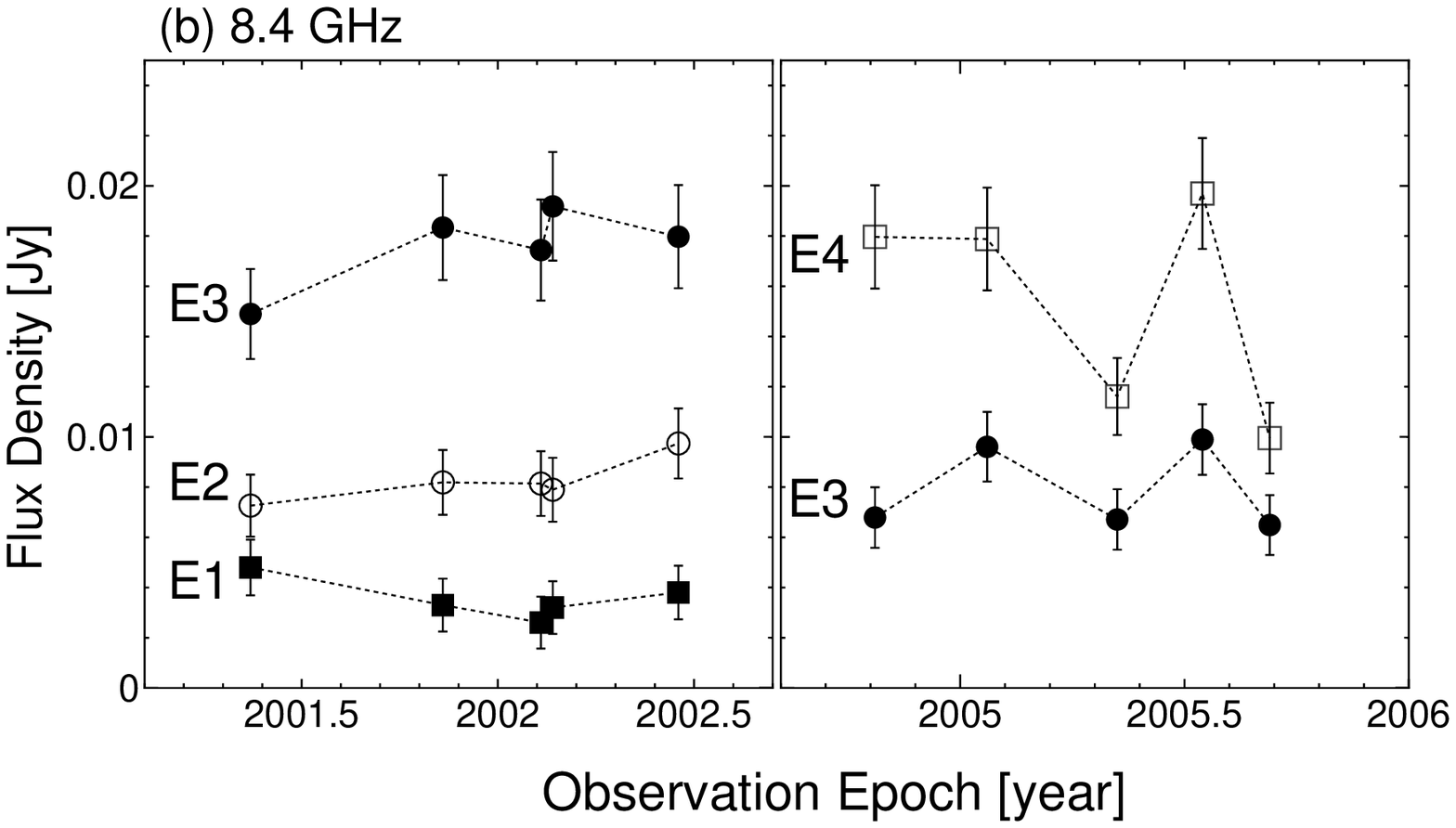}

  \caption{Flux variation of the jet components, (a) at 2.3 GHz, and (b) at 8.4 GHz. The components A at 2.3 GHz and E5 at 8.4 GHz are excluded in this figure, because of insufficient number of data points.
No significant systematic variation cannot be found at both frequencies. 
Note that the sudden flux dip at the 2005.35 epoch at both frequencies may be affected by the lower-resolution due to the lack of longer baseline data related to the St. Croix station. 
}
 \end{center}
\end{figure}


\begin{figure}
  \begin{center}
\epsscale{0.75}
 \plottwo{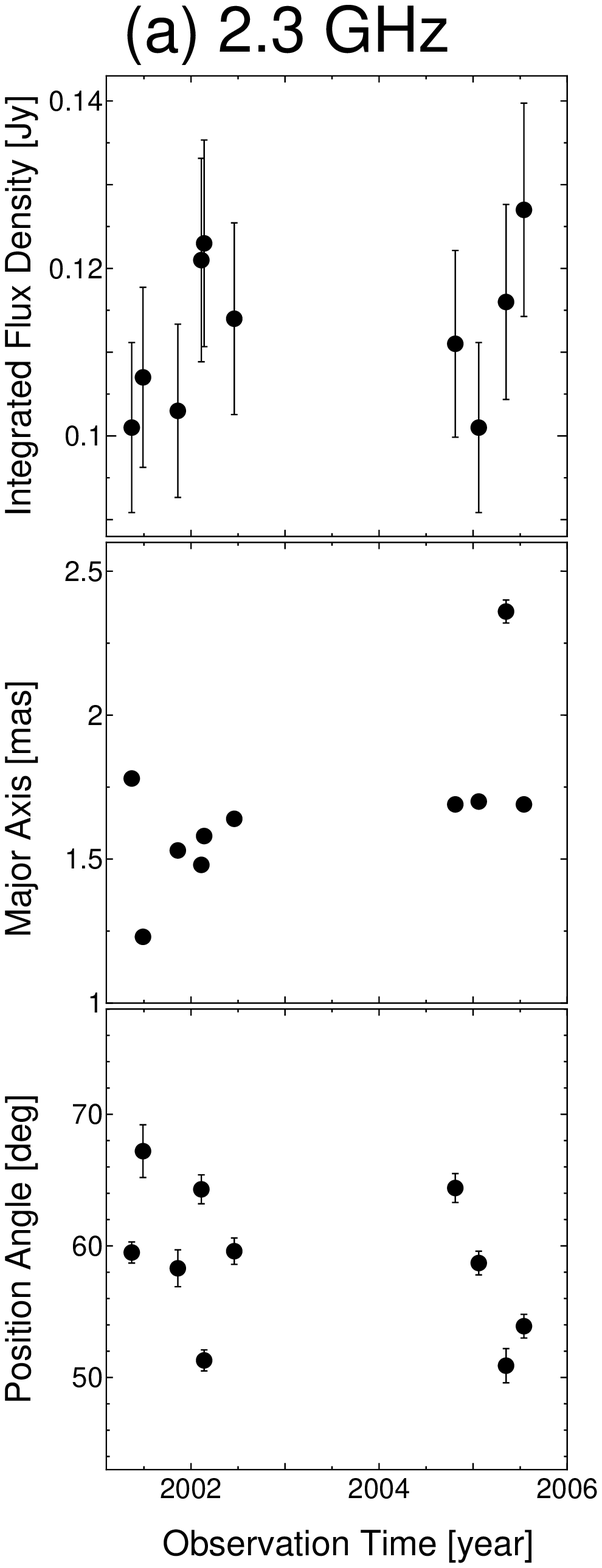} {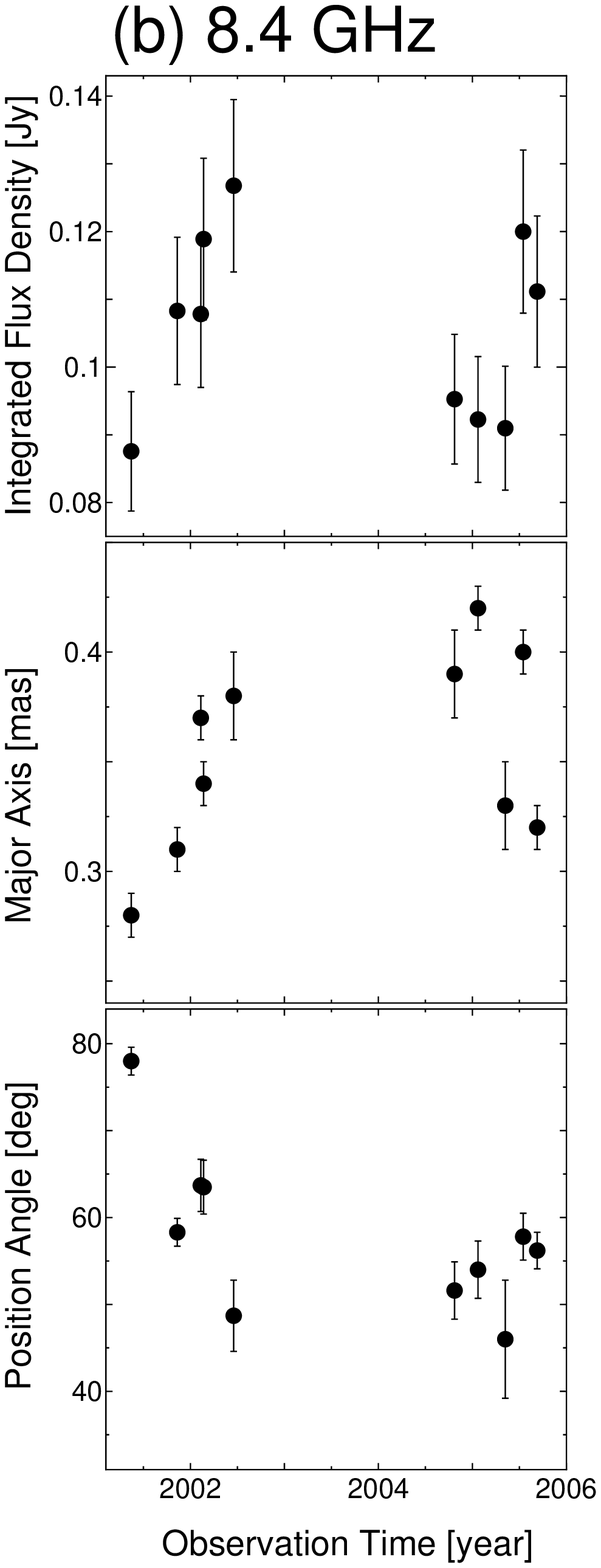}
  \caption{Time evolution of the fitted core parameter (a) at 2.3 GHz, and (b) at 8.4 GHz.  top : integrated flux density, middle : major axis, and bottom : position angle. The major axis increased with time from 0.28 to 0.38 mas in Period 1. After reaching its largest value of 0.42 mas, it decreased to 0.32 mas at the last epoch in Period 2.  
 The flux density also increased about 50 $\%$, and the PA changed from 80 to 50 degrees in Period 1, and a possible increase in the major axis and flux density can also be seen at 2.3 GHz.}
 \end{center}
\end{figure}


\begin{figure}
  \begin{center}
 \epsscale{0.75}
 \plotone{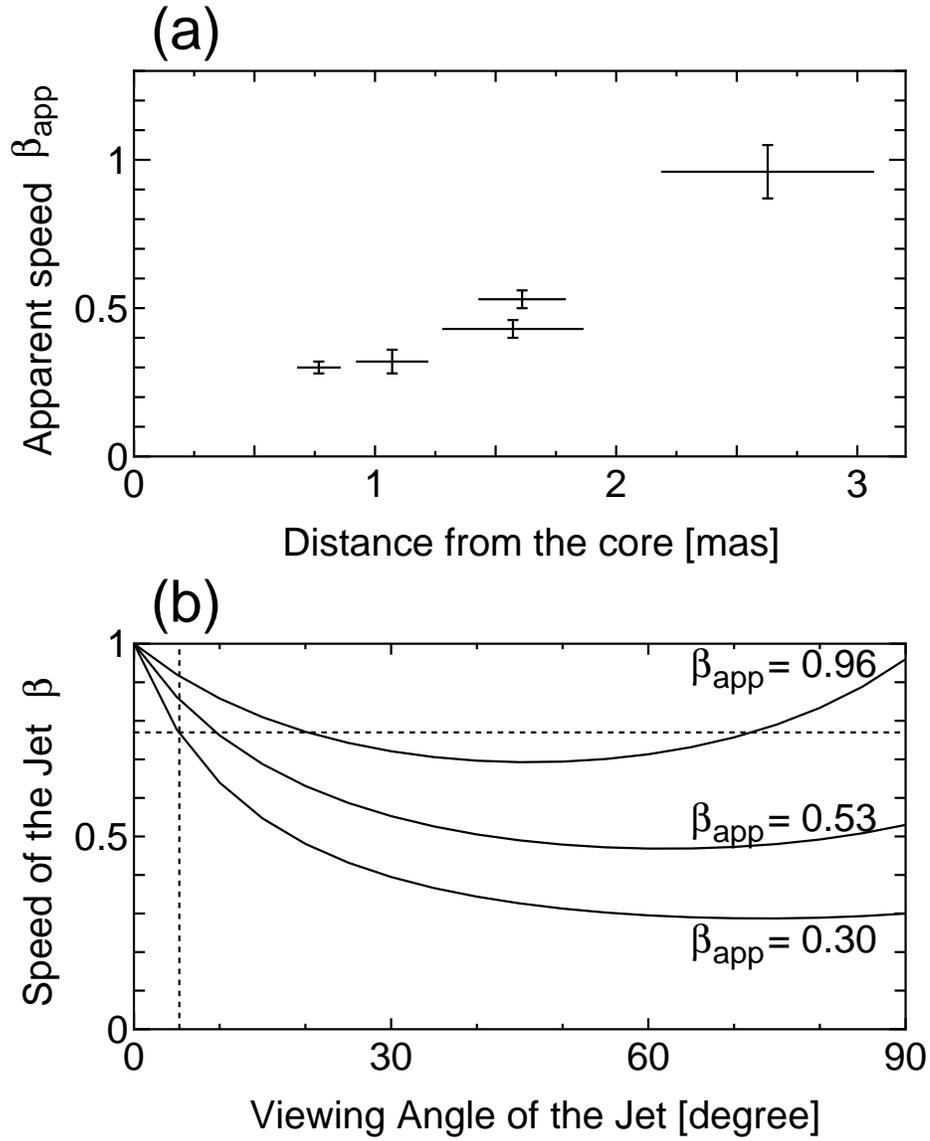}
  \caption{(a) Evolution of $\beta_{\rm app}$ as a function of distance from the core, (b) Constraints on the viewing angle and speed of a jet from  $\beta_{\rm app}$ in Period 1, according to the relativistic jet model. 
  The horizontal dotted line indicates a jet speed  of 0.77 $c$, and the vertical dotted line indicates a viewing angle of 5.3 degrees. Assuming that  the viewing angle is always 5.3 degrees,  we found that the jet accelerates from 0.77 $c$ to 0.92 $c$.
  Assuming that the jet speed is always  0.77 $c$, we found that the viewing angle changes from 5.3 to 20.8 degrees. 
}
 \end{center}
\end{figure}

\end{document}